\pdfoutput=1 
\newcommand*{\ATLASLATEXPATH}{latex/}
\documentclass[cernpreprint,texlive=2011,txfonts,UKenglish]{latex/atlasdoc} 
\usepackage{\ATLASLATEXPATH atlaspackage}
\usepackage{\ATLASLATEXPATH atlasphysics}
\usepackage{\ATLASLATEXPATH atlasbiblatex}
\usepackage{\ATLASLATEXPATH atlascontribute}
\usepackage{\ATLASLATEXPATH atlasphysics}
\usepackage{verbatim}
\graphicspath{{logos/}{figures/}}
\usepackage{paper2015-defs}

\AtlasTitle{Search  for strong gravity in multijet final states produced in pp collisions at $\sqrt{s}~=$~13 TeV using the ATLAS detector at the LHC }

\author{The ATLAS Collaboration}

\AtlasRefCode{EXOT-2015-09}

\usepackage{multirow}

\AtlasNote{}
\PreprintIdNumber{CERN-PH-EP-2015-312}

\AtlasAbstract{
  A search is conducted for new physics in multijet final states using 3.6 inverse femtobarns of data from proton-proton collisions at $\sqrt{s}=13$ \TeV\ taken at the CERN Large Hadron Collider with the ATLAS detector.
Events are selected containing at least three jets with scalar sum of jet transverse momenta ($H_\mathrm{T}$)  greater than 1 \TeV.
 No excess is seen at large $H_\mathrm{T}$ and limits  are presented on new physics: models which produce final states containing at least three jets and having  cross sections larger than 1.6 fb with  $H_\mathrm{T}>$ 5.8 \TeV\ are excluded. Limits are
also given  in terms of new physics models of strong gravity that hypothesize 
additional space-time dimensions. 
}

\AtlasJournal{JHEP}

\hypersetup{pdftitle={ATLAS draft},pdfauthor={The ATLAS Collaboration}}
\begin{document}
\author{The ATLAS Collaboration}
\maketitle

\section{Introduction}
\label{sec:intro}
Many models of gravity postulate a fundamental gravitational scale comparable to the  electroweak scale, hence allowing the production of non-perturbative
gravitational states, such as micro black holes and string balls
(highly excited string states) at Large Hadron Collider (LHC) collision
energies~\cite{ArkaniHamed:1998rs,Antoniadis:1998ig,Randall:1999ee,Randall:1999vf}.   
If black holes or string balls with masses much
higher than this fundamental gravitational scale  are produced  at the LHC, they behave as
classical thermal states and decay to a relatively large number of high
transverse momentum (\pT) particles.
One of the predictions of these models is that particles are
emitted from black holes at rates which primarily depend on the number of
 Standard Model (SM) degrees of freedom (number of charge, spin, flavour, and
colour states). 
Spin-dependent effects, such as the Fermi--Dirac and Bose--Einstein
distributions in statistical mechanics, and gravitational transmission
factors (also dependent on spin) are less important. Several searches 
were carried
out using data from Run-1 at the LHC at centre-of-mass energies 
of 7 and 8 $\TeV$ 
by ATLAS~\cite{Aad:2015zra,Aad:2011bw,Aad:2013lna,Aad:2012ic,Aad:2014gka} 
and CMS~\cite{Khachatryan:2010wx,Chatrchyan:2012taa,Chatrchyan:2013xvaa}. The analysis 
described here follows the method of a similar ATLAS analysis using  8  $\TeV$ data~\cite{Aad:2015zra}.
The increase in the LHC energy to 13 $\TeV$ in Run-2  brings a large increase in the sensitivity 
compared to Run-1; for the data set used here the increase is of the order of 50\% in the energy scale being probed.  Another analysis looking at dijet final states~\cite{ATLAS:2015nsi} is also sensitive to new physics of the type discussed here.

Identification of high-\pT, high-multiplicity final states resulting from the 
decay of high-mass objects is accomplished by studying  the scalar sum of the jet \pT (\HT). 
A low-\HT{} control region is defined. New physics of the type considered in this paper cannot contribute significantly in this region
as it is excluded by the previous searches.
A fit-based technique is used to extrapolate from the control region to a
high-\HT{} signal region to estimate the amount of the SM background.   
The observation is compared to the background-only expectation determined by the fit-based method.
In the absence of significant deviations from the background-only
expectation, 95\% Confidence Level (CL) limits on micro black hole and string-ball production are  set. The limits are given in terms of parameters
 used in the \charybdis~1.0.4~\cite{Frost:2009cf} model.

The production and decay of black holes and string balls lead to final
states distinguished by a high multiplicity of high-\pT{} particles, 
consisting mostly of jets arising from quark and gluon emission.
Since black-hole decay is  considered to be a stochastic process, 
different numbers of particles, and consequently  jets, are  emitted from black holes
with identical kinematic distributions.
This motivates the search in inclusive jet multiplicity slices, rather than
optimizing a potential signal-to-background for a particular exclusive jet
multiplicity.

The analysis is not optimized for any particular model. 
However, for the purpose of comparison to other searches both within ATLAS and between the LHC experiments, 
\charybdis~1.0.4 is used. For the micro black holes 
 the number of extra dimensions in the model is fixed to be two, four or six,  the black hole is required to be rotating, and the limits are
presented as a function of the fundamental Planck scale (\MD) and the mass threshold (\Mth). In the case of string balls, limits are presented
as functions of \Mth, the string scale (\Ms) and the string coupling (\gs).

\section{ATLAS detector}
\label{sec:detector}

The ATLAS detector~\cite{Aad:2008zzm} covers almost the whole solid angle around the collision point with layers of 
tracking detectors, calorimeters and muon chambers. For the measurements presented in this note, 
the calorimeters 
are of particular importance.
The inner detector, immersed in a magnetic field provided by a solenoidal magnet,
 has full coverage in  $\phi$ and covers the pseudorapidity\footnote{ATLAS uses
a right-handed coordinate system with its origin at the nominal
interaction point in the centre of the detector and the $z$-axis along
the beam direction.
The $x$-axis points toward the centre of the LHC ring, and the $y$-axis
points upward. 
Cylindrical coordinates $(r,\phi)$ are used in the transverse
plane, $\phi$ being the azimuthal angle around the  $z$-axis.
The pseudorapidity $\eta$ is defined in terms of the polar angle
$\theta$ by $\eta \equiv -\ln[\tan(\theta/2)]$.}  range $|\eta|<2.5$.
It consists of a silicon pixel detector, a silicon microstrip detector and 
a transition radiation straw-tube tracker. The innermost pixel layer, the insertable B-layer, was added 
between Run-1 and Run-2 of the LHC,  at a radius of 33 mm around a new, thinner, beam pipe. 
In the pseudorapidity region $|\eta| < 3.2$, high granularity
liquid-argon  (LAr) electromagnetic (EM) sampling calorimeters are
used.
An iron-scintillator tile calorimeter provides hadronic coverage
over $|\eta| < 1.7$.
The end-cap and forward regions, spanning $\HTmin < |\eta| < 4.9$, are
instrumented with LAr calorimetry for both EM and hadronic
measurements.
The muon spectrometer surrounds these calorimeters, and comprises 
 a system of precision tracking chambers for muon reconstruction up to $|\eta|$ = 2.7 and trigger detectors with three large toroids,
each consisting of eight coils providing magnetic fields for the muon detectors.

\section{Event selection}
\label{sec:selection}

The data used here were recorded in 2015, with the LHC
operating at a centre-of-mass energy of $\sqrt{s} = 13\TeV$. 
All detector elements are required to be fully operational, except for the insertable B-layer of pixels which was not operating for a small subset of the data. Data corresponding to a total
integrated luminosity of \lumi{} are used in this analysis measured with an
 uncertainty of 9\%.
This  is derived following the same methodology as that detailed in
Ref.~\cite{Aad:2013ucp},
from a preliminary calibration of the luminosity scale using a pair of $x$--$y$ beam-separation scans performed in June 2015.

The ATLAS detector has a two-level trigger system: the level-1 hardware stage and the high-level
trigger software stage.
The events used in this search are selected using a high-\HT{} trigger, 
which requires at least one jet of hadrons with $\pT > 200\GeV$, 
and a high scalar sum of transverse momentum of all the jets in the events, $\HT{} > 0.85\TeV$.
In this analysis, a requirement of $H_{\mathrm{T}}~>~$1 $\TeV$ is applied, for which the trigger is fully efficient. 
All jets included in the computation of $\HT{}$ are required to satisfy
 $\pT > 50\GeV$ and $|\eta|<2.8$.

Events are required to have a primary vertex with at least two
associated tracks with \pT{} above $0.4\GeV$. 
The primary vertex assigned to the hard-scattering collision is the one
with the highest $\sum_\mathrm{track} \pT^2$, where the scalar sum of
track $\pT^2$ is taken over all tracks associated with that vertex. 

Since black holes and string balls are expected to decay predominantly
to quarks and gluons, the search is simplified by considering only jets.  
The analysis uses jets of hadrons, as well as misidentified jets arising from
photons, electrons, and taus. Using the hadronic energy calibration instead of the dedicated
calibration developed for these objects leads to small energy shifts. Since particles of these types are expected to occur in
less than 0.6\%  of the signal events in
the data sample (as determined from simulation studies), such calibration effects do not contribute significantly to the resolution
of $H_{\mathrm{T}}$.   

The anti-$k_t$ algorithm~\cite{Cacciari:2008gp} is used for jet clustering, with
a radius parameter $R = 0.4$.
The inputs to the jet reconstruction are three-dimensional
clusters comprised of energy deposits in the calorimeters~\cite{Lampl}.
This method first clusters together topologically connected calorimeter
cells and then classifies these clusters as either electromagnetic or
hadronic.  
The four-momenta corresponding to these clusters are calibrated for the response to incident hadrons using the procedures
described in Refs.~\cite{Aad:2015ina,ATLAS-CONF-2015-002}. The agreement between data and simulation is further improved by the 
application of a residual correction derived in situ at lower collision energies~\cite{ATLAS-CONF-2015-017}
which was  validated for use at 13 $\TeV$ through additional extrapolation uncertainties~\cite{ATL-PHYS-PUB-2015-015}.

While a data-driven method is used to estimate the background, 
simulated events are used to establish, test and validate the methodology of the analysis. 
Therefore, simulation is not required to accurately describe the background, but it should be 
sufficiently similar that the strategy can be tested before applying it to data.
Multijet events constitute the dominant background in the search region, with small contributions from top-quark pair-production  ($\ttbar$), $\gamma$ + jets; 
\Wboson + jets,~\Zboson + jets, single-top quark, and diboson background contributions are negligible.

The baseline multijet sample of inclusive jets is generated using \pythia~8.186~\cite{Pythia:2007} implementing leading order (LO) perturbative QCD matrix elements with
NNPDF23\_ lo\_ as\_ 0130\_ qed parton distribution functions (PDFs)~\cite{Ball:2014uwa}
 for $2\rightarrow 2$ processes and \pT-ordered parton showers
calculated in a leading-logarithmic approximation with the ATLAS A14  set of tuned parameters (tune)~\cite{atlas:pythia8tunes}.
A reasonable agreement in the shape of the $H_{\mathrm{T}}$ distribution was observed in Run-1 for different inclusive jet multiplicity
 categories~\cite{Aad:2015zra}. 
All Monte-Carlo (MC) simulated background samples are passed
through a full GEANT4~\cite{Agostinelli2003250} simulation of the ATLAS
detector~\cite{ATLASSimInfra}.
Signal samples are generated from the  \charybdis~1.0.4 MC event generator, 
which is run with leading-order PDF CTEQ6L1~\cite{Martin:2009iq} and uses the \pythia~8.210
generator for fragmentation with the ATLAS A14 tune.
The most important parameters that have significant effects on micro black hole production
are the number of extra dimensions, the (4+n)-dimensional Planck scale \MD\ and the black hole mass threshold \Mth.
Signal samples are generated on a grid of \MD\ and \Mth\ for $n=2$, 4 and 6. In the case of string-ball production
two sets of samples are produced; one as a function of \Mth\  and the string scale \Ms\  for fixed value \gs~=~0.6 of the string coupling, and one as a function of \gs\ and \Mth\ for \Ms~= 3 $\TeV$.
The signal samples are passed through a fast detector simulation AtlFast-II~\cite{AFII}.
All simulated signal and background samples are reconstructed using the same software as the data.

\section{Analysis strategy}
\label{sec:strategy}

The search is performed by examining the \HT{} distribution for several inclusive 
jet multiplicities. For each multiplicity, three regions of \HT{} are used: control ($C < \HT{} < V$), validation  ($V < \HT{} < S$)
and signal ($\HT{}>~S$).
Data in the control region are fitted to an empirical function which is then extrapolated to
 predict the event rate
in the validation and signal regions in the absence of new physics. 
The boundaries of these regions ($C$, $V$ and $S$) depend on the integrated luminosity of the data sample
 used and inclusive jet multiplicity requirement. The following criteria must be satisfied:
the lower boundary of the control region ($C$) should be sufficiently large that the shape of the  \HT{}
 distribution near the boundary is not distorted by event selection effects; contamination from a possible
 signal due to  new physics in the 
control region must be small for all possible signals not excluded by prior results.
There should be 
some background events in the validation region whose lower boundary is deterined by $V$,  with a small  signal to background ratio from 
signals that are not excluded by a previous analysis,
so that the background extrapolation can be checked.  
The signal region is defined so that the background extrapolation uncertainty relative to the background prediction is small: the boundary $S$ is chosen so that the
 (pseudo-experiment-based,  see below) background extrapolation uncertainty is approximately 0.5 
 events for $H_{\mathrm{T}}~\ge~S$.

A large increase in sensitivity to new physics is expected in Run-2  primarily due
to the increase in centre-of-mass energy.
A data set of a few  \ifb\ has such a large range
of sensitivity that significant signal 
contamination in the control and validation regions is possible. Therefore,
 a bootstrap approach is adopted and 
data sets are examined whose size increases by approximately a factor of ten at each step, 
starting with a sample whose
sensitivity is slightly beyond the Run-1 limit; 
simulation studies indicate an initial integrated luminosity of up to 10 \ipb~
 would be free of signal contamination.  This
 will ensure that if a search in one
step  sees no new physics, the possible contributions of signal to the control 
and validation regions of the next
step  are small.  
For each data set the boundaries of the regions
are determined  as follows. Simulations    are  normalized to data
in the normalization region, which is 1.5 \TeV < $H_{\mathrm{T}}$ < 2.9 \TeV. First, the lower boundary of the validation region,
$V$, is chosen from this normalized MC simulation so that at least 20 events are expected for $S~>$~\HT{}$~>~V$. This will allow a quantitative comparison of the data and expectation in the
validation region to check that the extrapolation procedure is working properly. The lower boundary of the control region ($C$) is determined by requiring that 
 the fit functions applied to MC-pseudo-data have a reduced $\chi^2$ 
distribution peaked near one and then choosing $C$ to minimize the pseudo-experiment-based uncertainty. Finally, the lower boundary of  the signal  region,
$S$, is chosen so that the extrapolation uncertainty is approximately 0.5 events for $H_{\mathrm{T}}~>~S$.

The total data set used corresponds to an integrated luminosity of \lumi. 
A four-step bootstrap is adopted using exclusive data sets, for which \lumia~is used in the first step,
 \lumib~in the second, \lumic~in the third and the remaining \lumid~is used in the last step.
The observed  \HT{} distribution is shown in Figure~\ref{fig:ht-6.6} for \lumia.
A comparison is made with MC simulation  for illustration. The MC simulation was normalized to the data in the normalization region independently in each jet multiplicity ($\njet$) sample.  
Lines delimiting the control, validation and signal regions are shown.
Before normalization the ratio Data/MC is approximately 0.74.
The example signal (\MD~=~2.5~$\TeV$, \Mth~=~6.0~$\TeV$) shown is just beyond the limit obtained from the Run-1 8 \TeV\ analysis. 
Any possible signal must therefore be smaller than this.  
 The \HT{} distribution expected from this signal is such that any 
 contamination in the control and validation  regions  is negligible.
 In addition, the contamination in the control region is negligible for all signals
 that this data set  (\lumia) is sensitive to. Possible contamination in the validation region is less than 10\% for signals not excluded by the Run-1 analysis.
 It can be seen from Figure~\ref{fig:ht-6.6} that data sets with high jet multiplicity contain rather few events. This first-step
 analysis therefore uses only the data sample with jet multiplicity, $\njet \ge 3$.

 The observed \HT{} distribution from the \lumib~sample used in the second step 
is shown in Figure~\ref{fig:ht-100} where comparison is made with MC simulation for illustration. 
The MC simulation  was  normalized to the data in the normalization region independently in each $n_{\mathrm{jet}}$ sample.
 Before normalization the ratio data/MC increases with jet multiplicity from 0.74 to 0.87.
This variation  is not unexpected since the MC simulation is leading order in QCD.
 Signal samples (\MD~=~3~\TeV, \Mth~=~7.5~\TeV) are superimposed on data in Figure~\ref{fig:ht-100}  which 
correspond approximately to those just beyond the sensitivity of the first-step analysis.
The logic of the previous paragraph applied here shows that the bootstrap approach is protected against signal contamination if data sets
increasing by a factor of ten in integrated luminosity are used.

The observed \HT{} distribution from the \lumic~sample used in the third step 
is shown in Figure~\ref{fig:ht-1000} where comparison is made with MC  simulation for illustration. 
The MC  simulation was normalized to the data in the normalization region independently in each $n_{\mathrm{jet}}$ sample. Before normalization the ratio Data/MC increases with jet multiplicity from 0.78 to 0.84.
Signal samples (\MD~=~4.5~\TeV, \Mth~=~8~\TeV) are superimposed on data in Figure~\ref{fig:ht-1000}  which 
correspond approximately to those just beyond the sensitivity of the second-step analysis.

Finally the observed \HT{} distribution from the \lumid\ sample used in the fourth step 
is shown in Figure~\ref{fig:ht-10000}, again  comparison is made with MC simulation for illustration. 
The MC simulation was\ normalized to the data in the normalization region independently in each $n_{\mathrm{jet}}$ sample. Before normalization the ratio Data/MC increases with jet multiplicity from 0.73 to 0.77. 
The ratio Data/MC is found to be consistent for all four exclusive data samples within statistical and luminosity uncertainties. 
 Signal samples (\MD~=~2.5~\TeV, \Mth~=~9.0~\TeV) are superimposed on data in Figure~\ref{fig:ht-10000}  which 
correspond approximately to those just beyond the sensitivity of the third-step analysis.

\begin{figure*}
\begin{center}                                   
\includegraphics[width=0.48\textwidth]{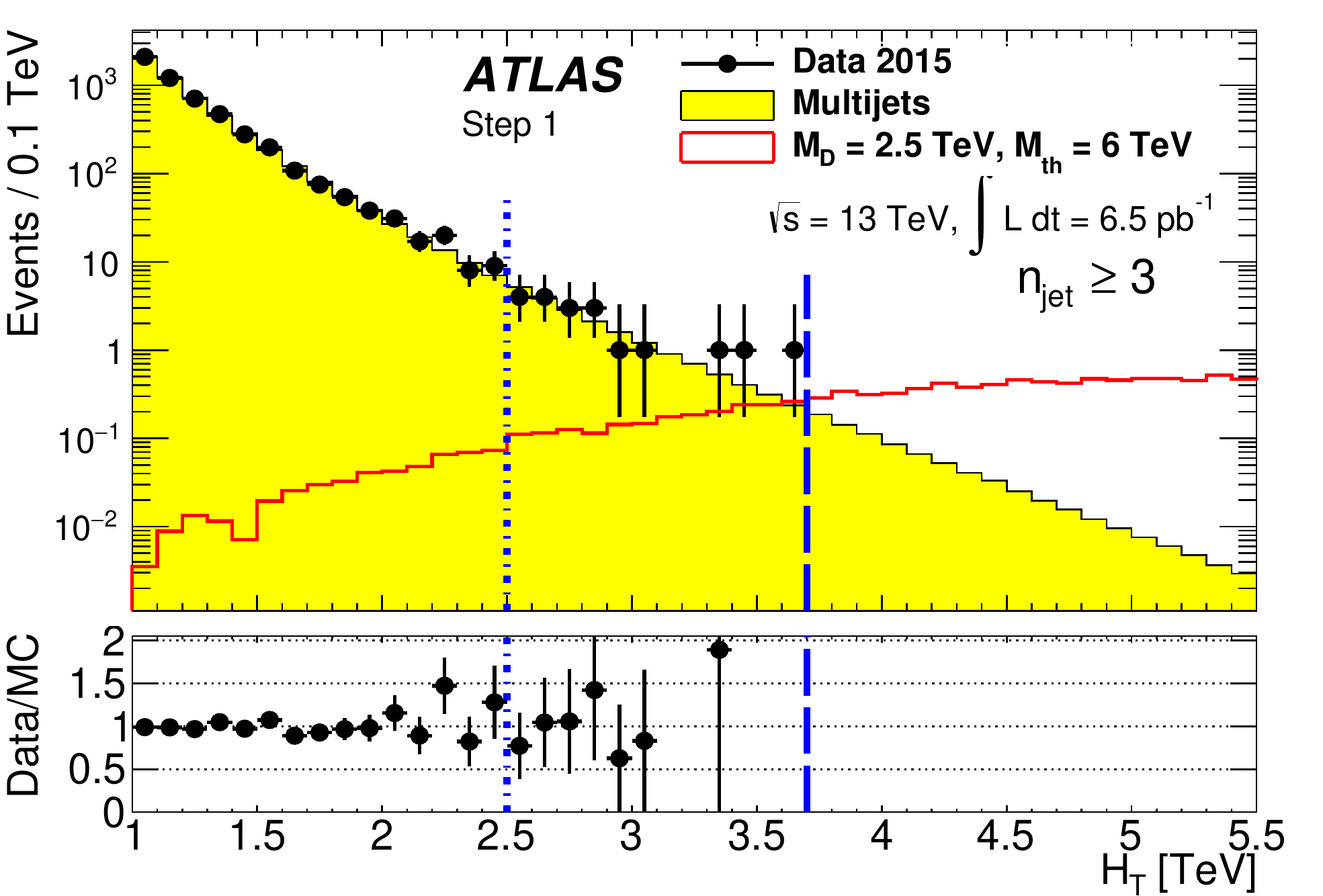}
\includegraphics[width=0.48\textwidth]{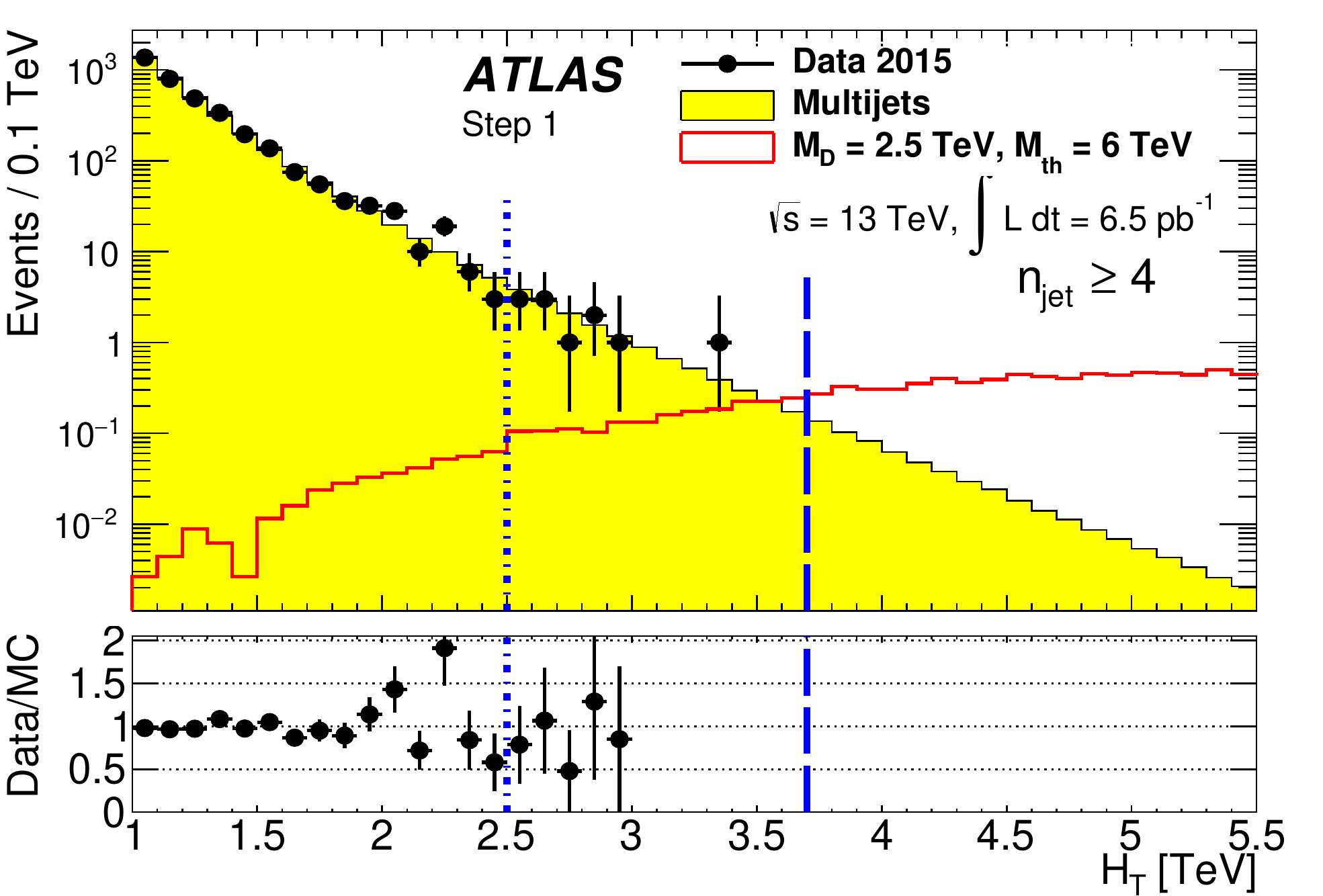}\\
\includegraphics[width=0.48\textwidth]{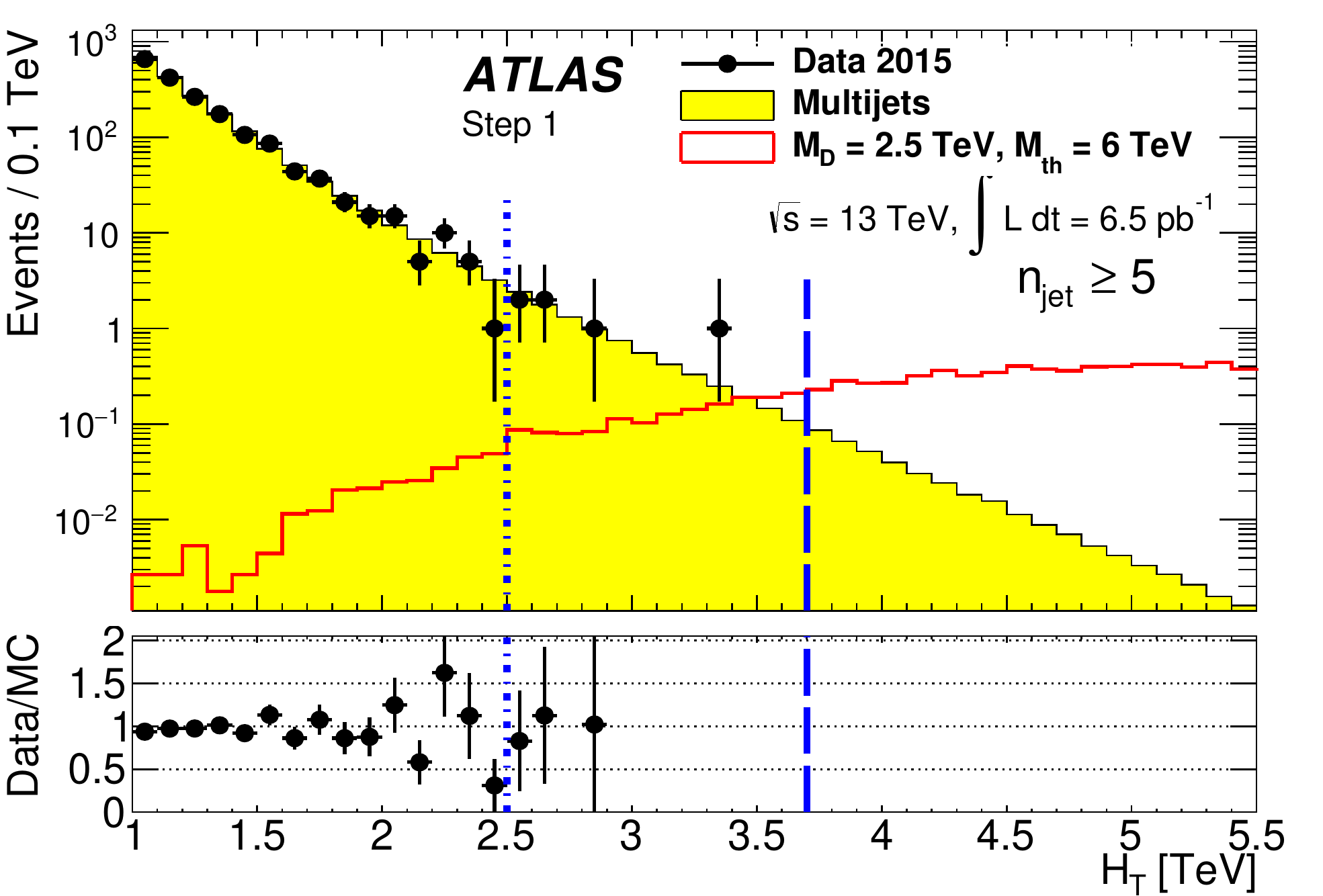}
\includegraphics[width=0.48\textwidth]{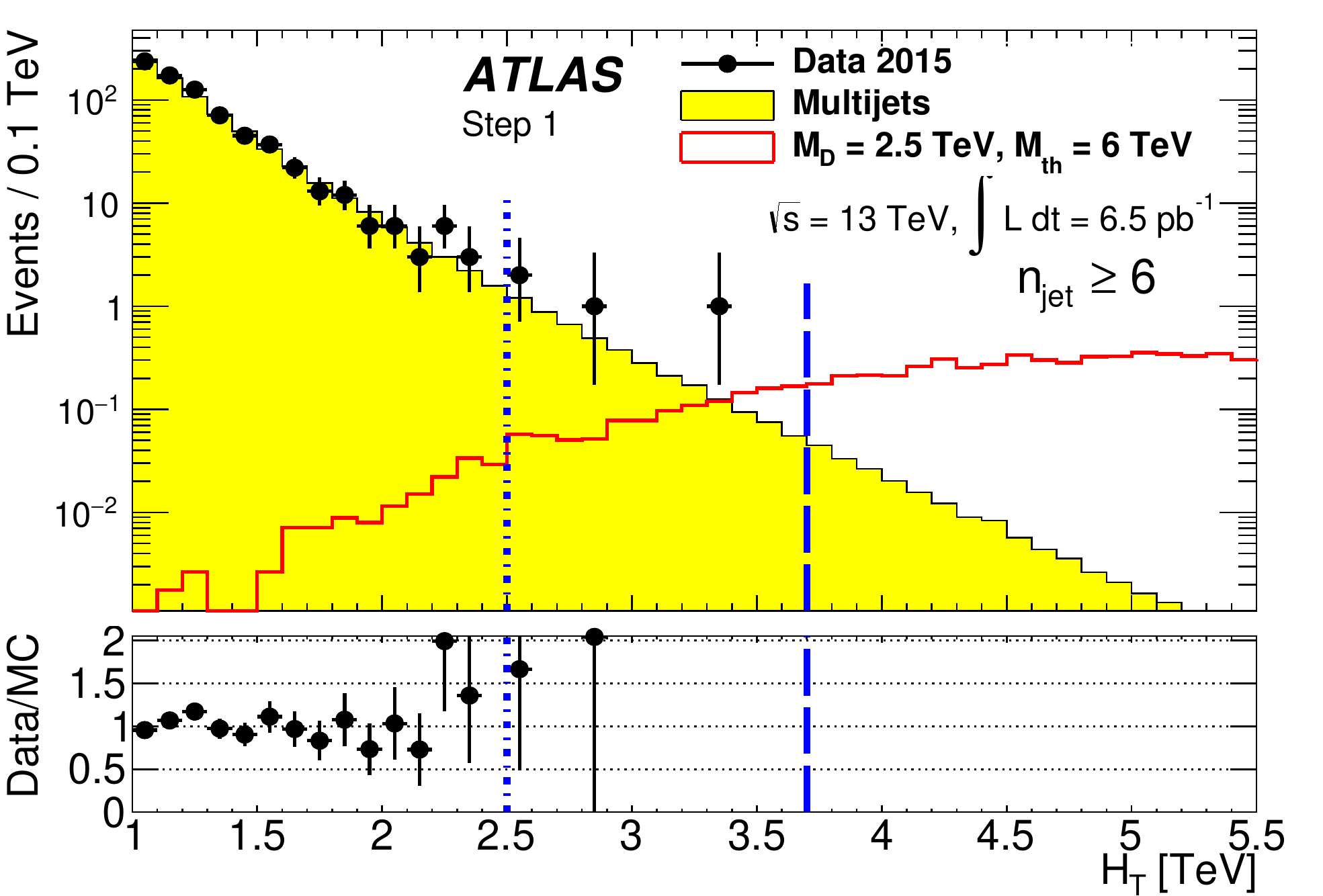}\\
\includegraphics[width=0.48\textwidth]{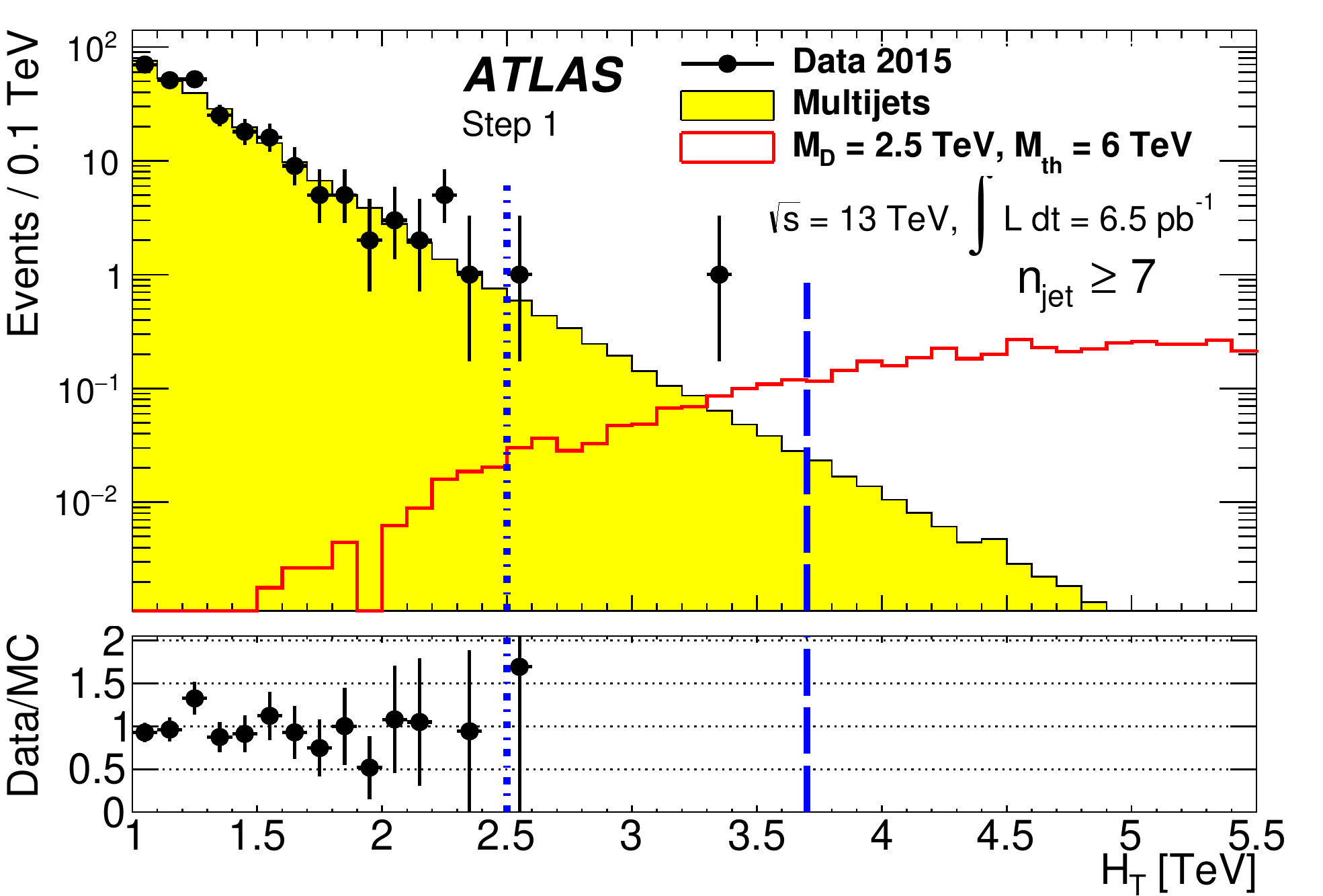}
\includegraphics[width=0.48\textwidth]{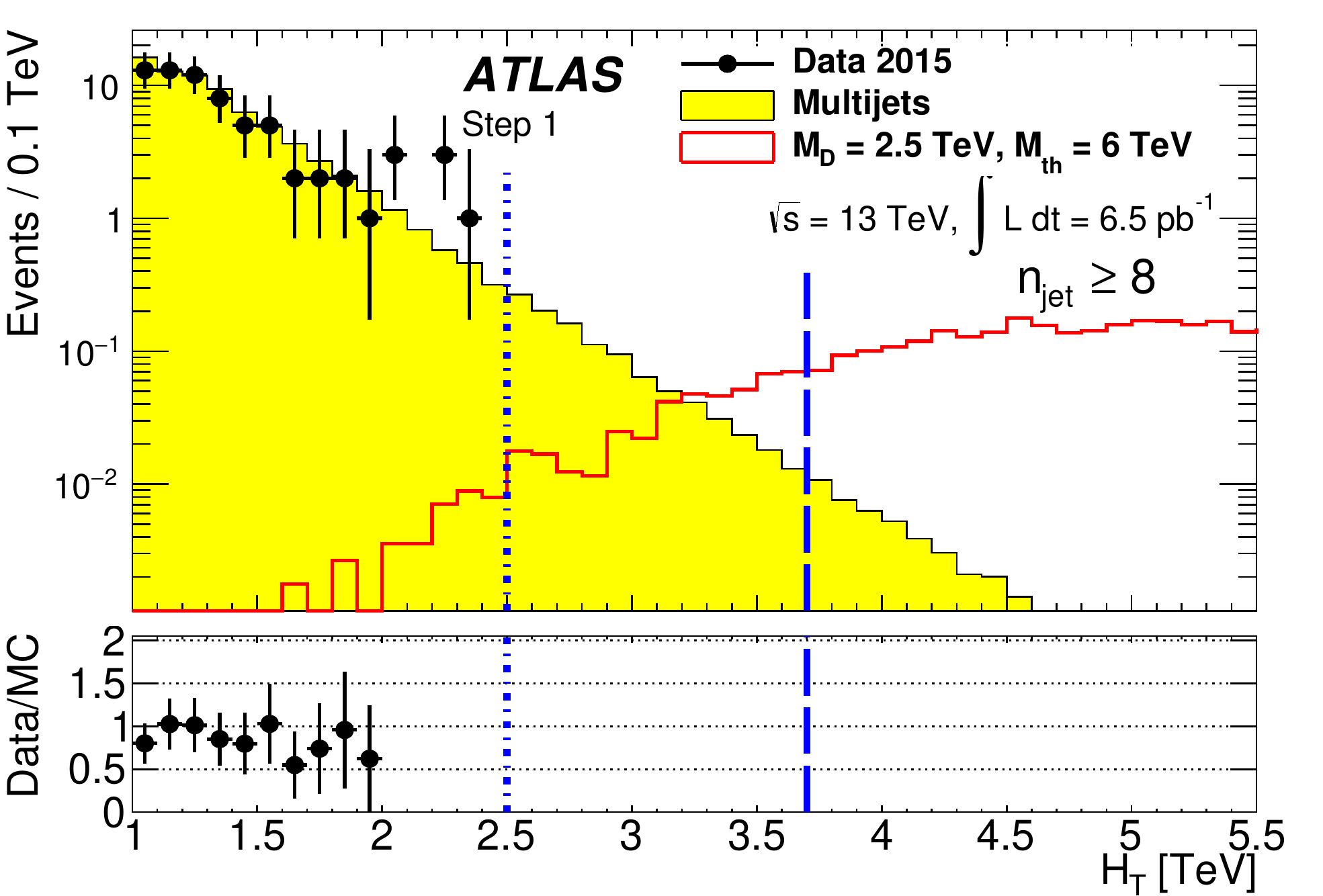}\\

\caption{Data and MC  simulation comparison for  the distributions of
the scalar sum of jet transverse momenta \HT{}  in different inclusive $\njet$ bins for  the \lumia~data sample.
The black hole signal with \MD~=~2.5~\TeV, \Mth~=~6.0~\TeV~is superimposed with the data and background MC simulation sample. The MC is normalized to data in the normalization region.
The vertical dashed-dotted line marks the boundary between control region and validation region, 
and the dashed line marks the boundary between validation region and signal region. The boundaries shown  correspond to those determined for the $\njet \ge 3$ case. 
}
\label{fig:ht-6.6}
\end{center}
\end{figure*}

\begin{figure*}
\begin{center}
\includegraphics[width=0.48\textwidth]{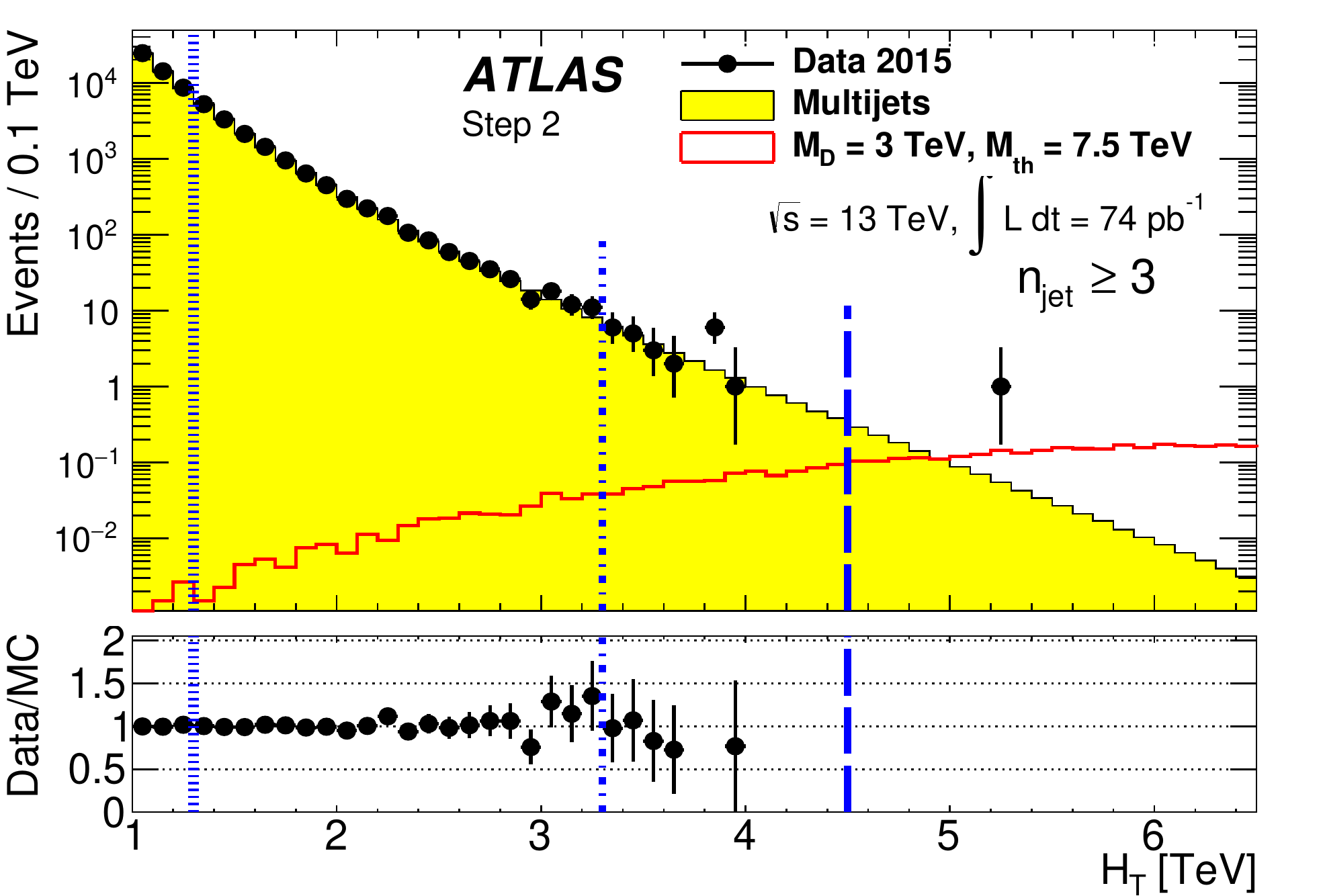}
\includegraphics[width=0.48\textwidth]{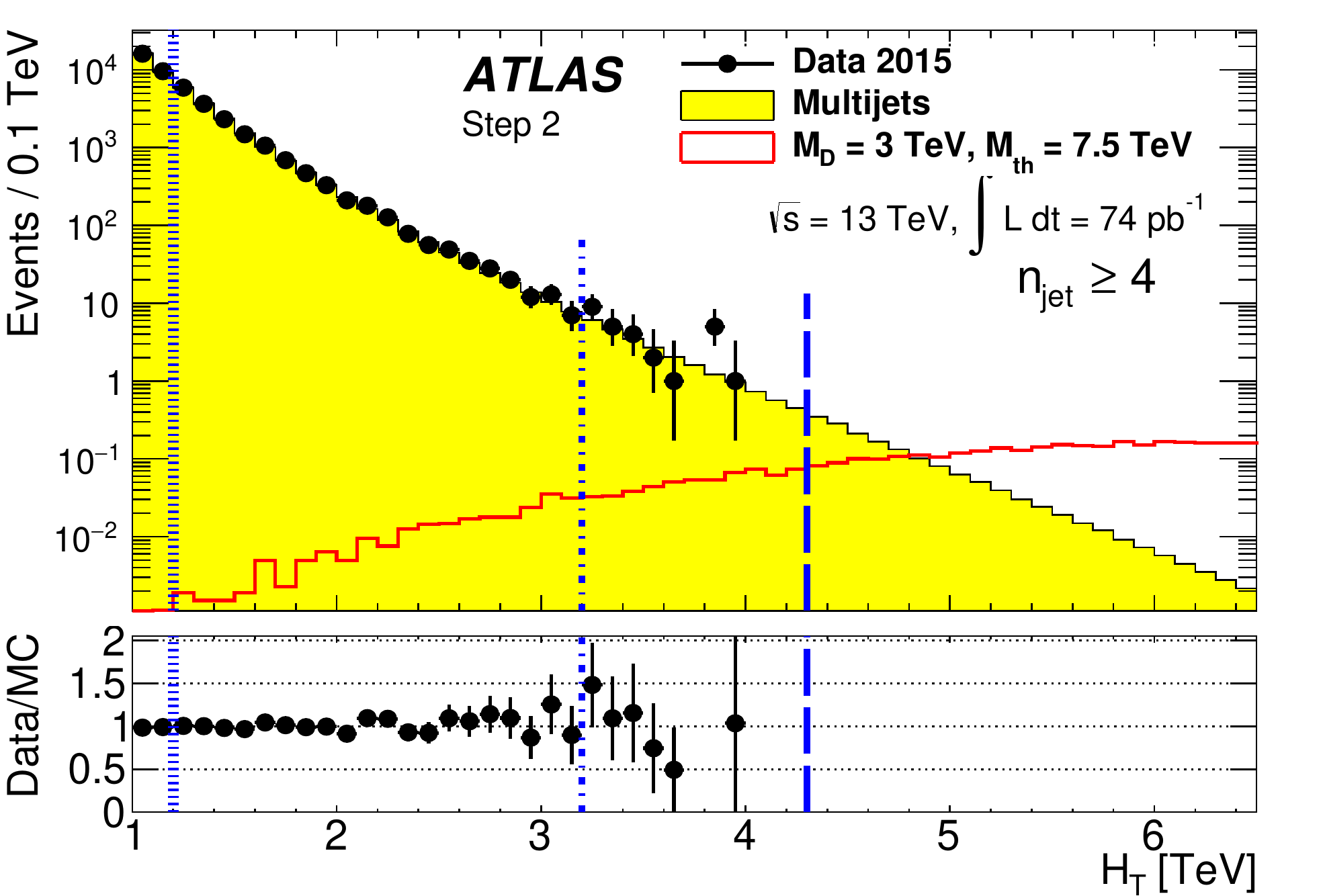}\\
\includegraphics[width=0.48\textwidth]{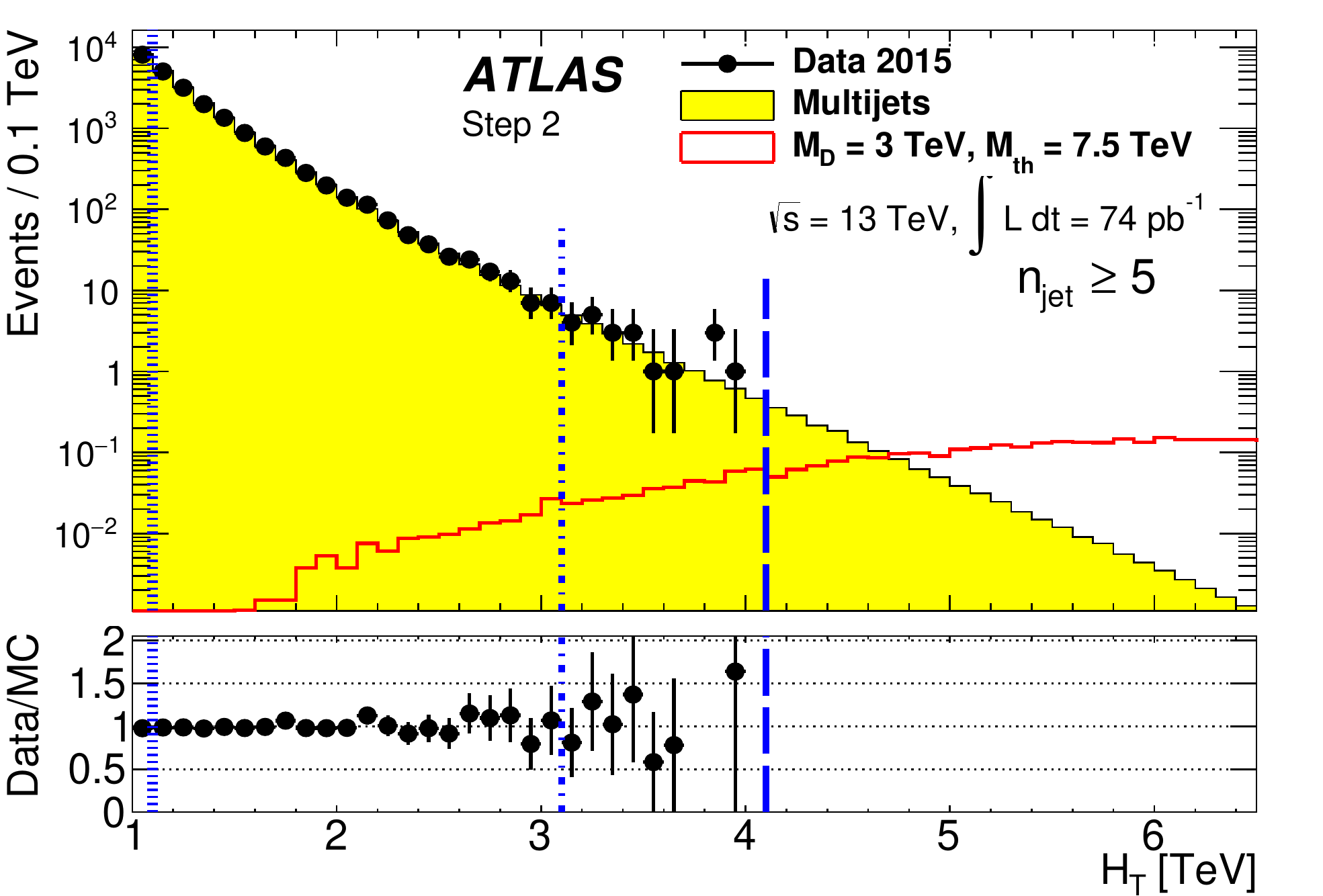}
\includegraphics[width=0.48\textwidth]{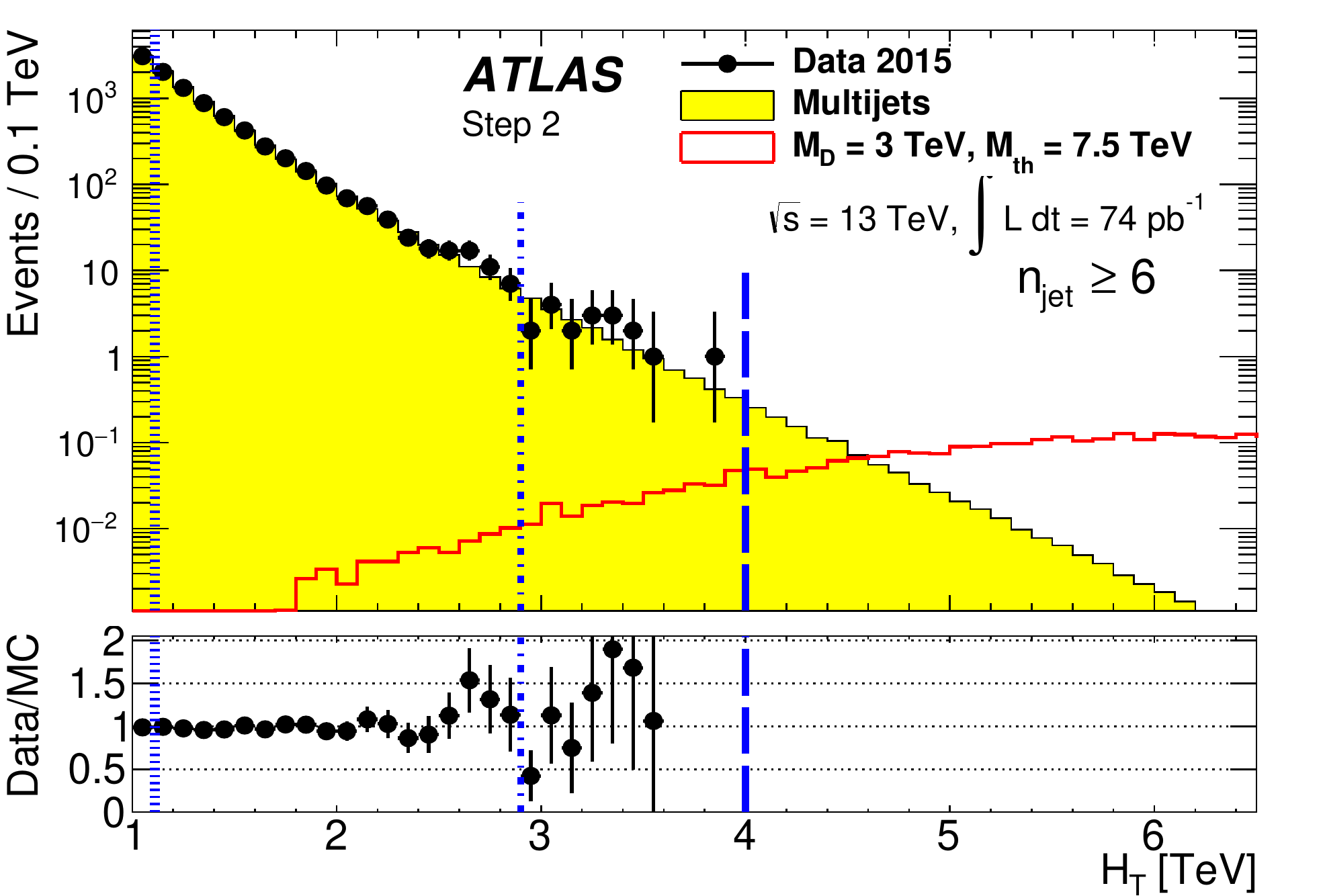}\\
\includegraphics[width=0.48\textwidth]{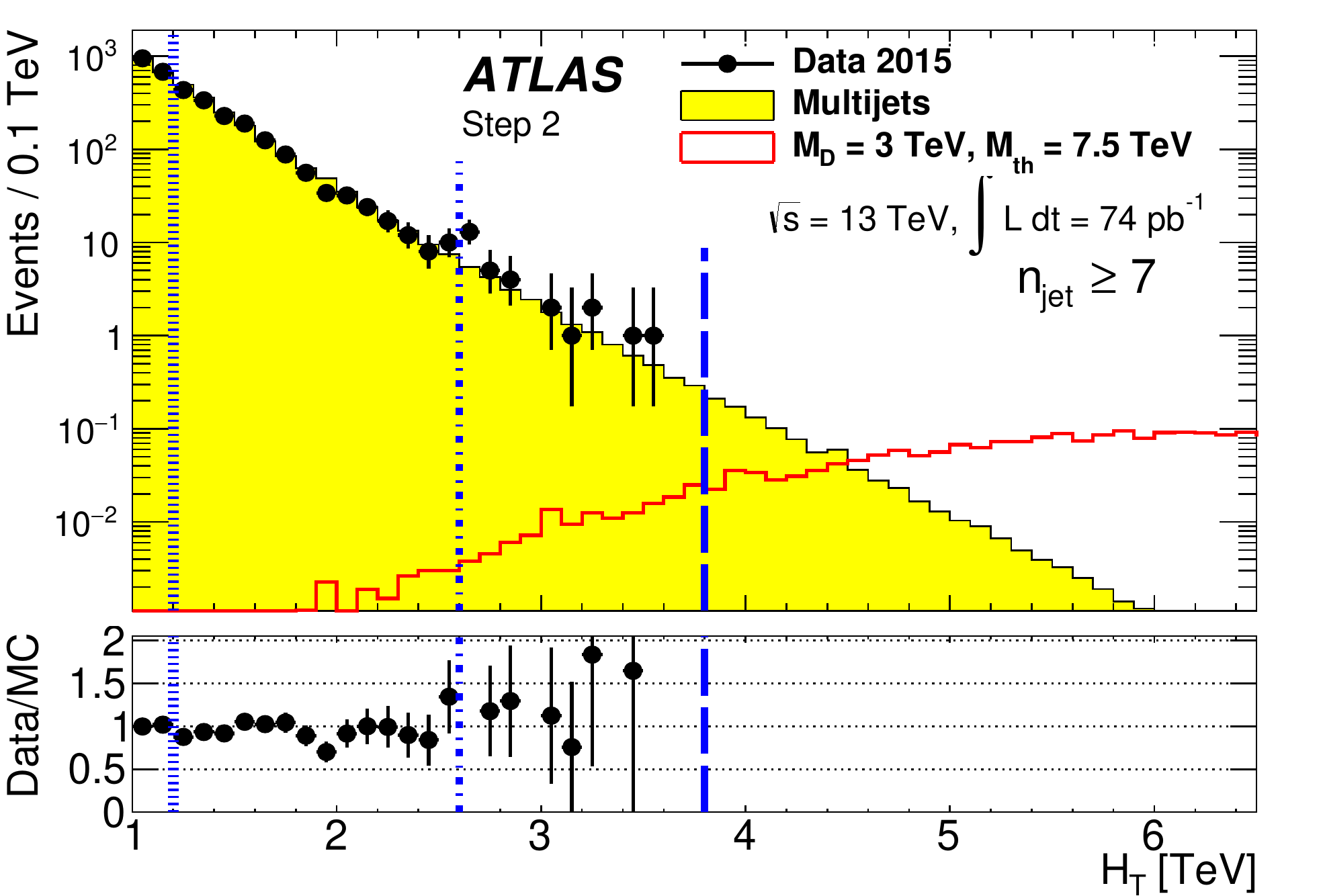}
\includegraphics[width=0.48\textwidth]{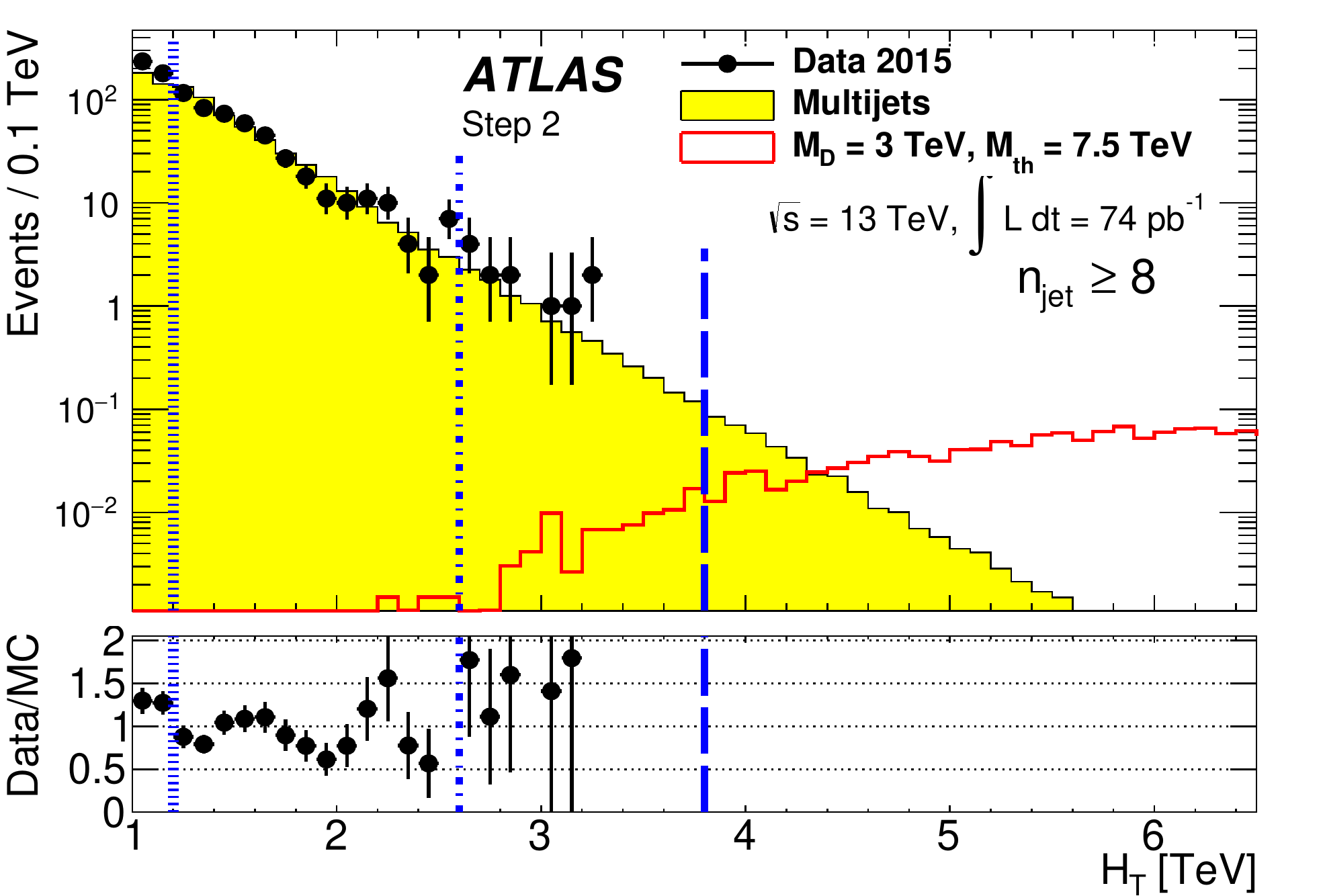}\\
\caption{ Data and MC simulation comparison for \HT{} distributions in different inclusive $\njet$ bins for the \lumib~data  sample.
The black hole signal with \MD~=~3~\TeV, \Mth~=~7.5~\TeV~is superimposed with the data and background MC.  The MC simulation was\ normalized to data in the normalization region.
The vertical dotted line marks the lower boundary of the control region, the vertical dashed-dotted line marks the boundary between control region and validation region, 
and the vertical dashed line marks the boundary between validation region and signal region. These boundaries are determined for each $n_{\mathrm{jet}}$ sample separately.
 }
\label{fig:ht-100}
\end{center}
\end{figure*}

\begin{figure*}
\begin{center}
\includegraphics[width=0.48\textwidth]{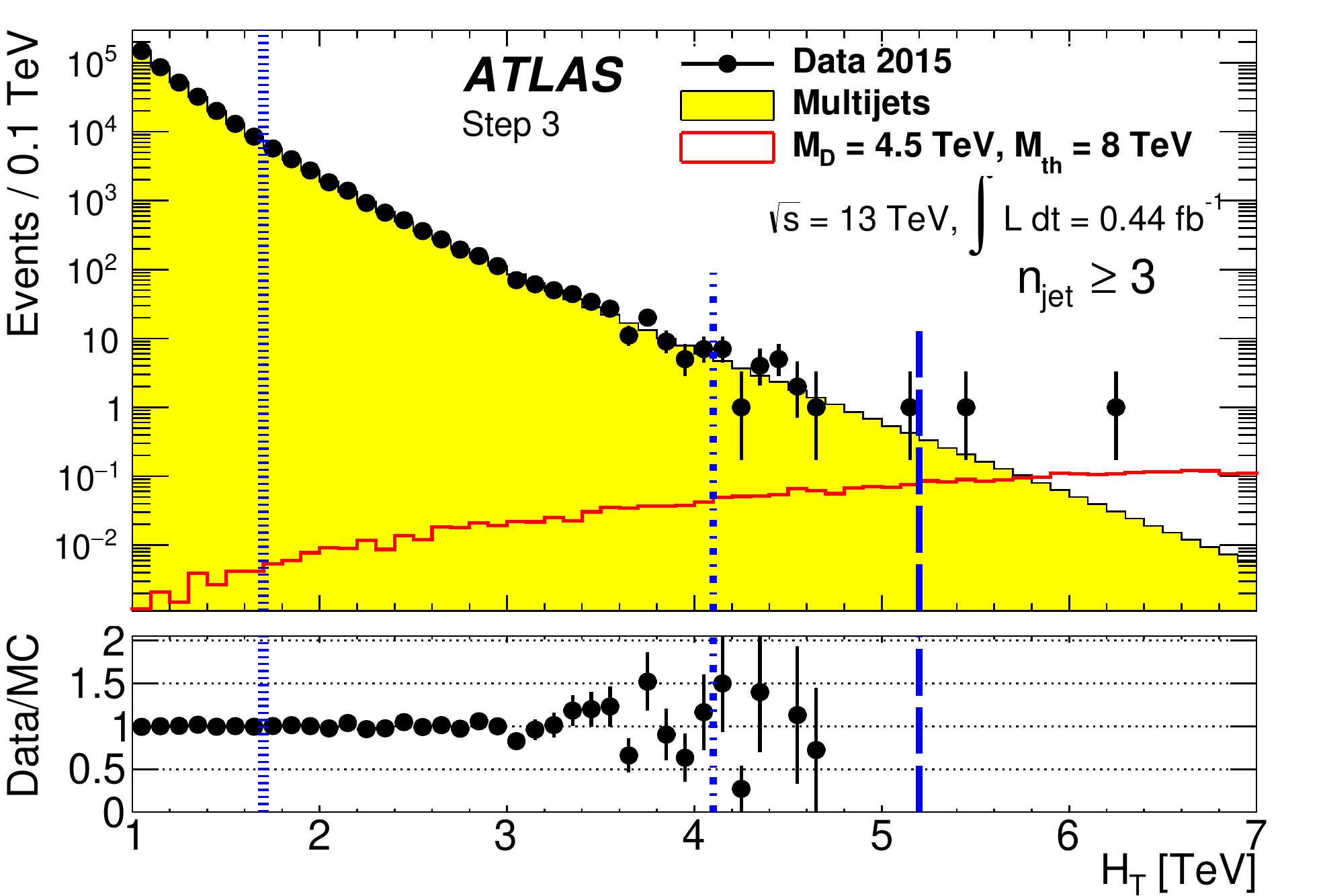}
\includegraphics[width=0.48\textwidth]{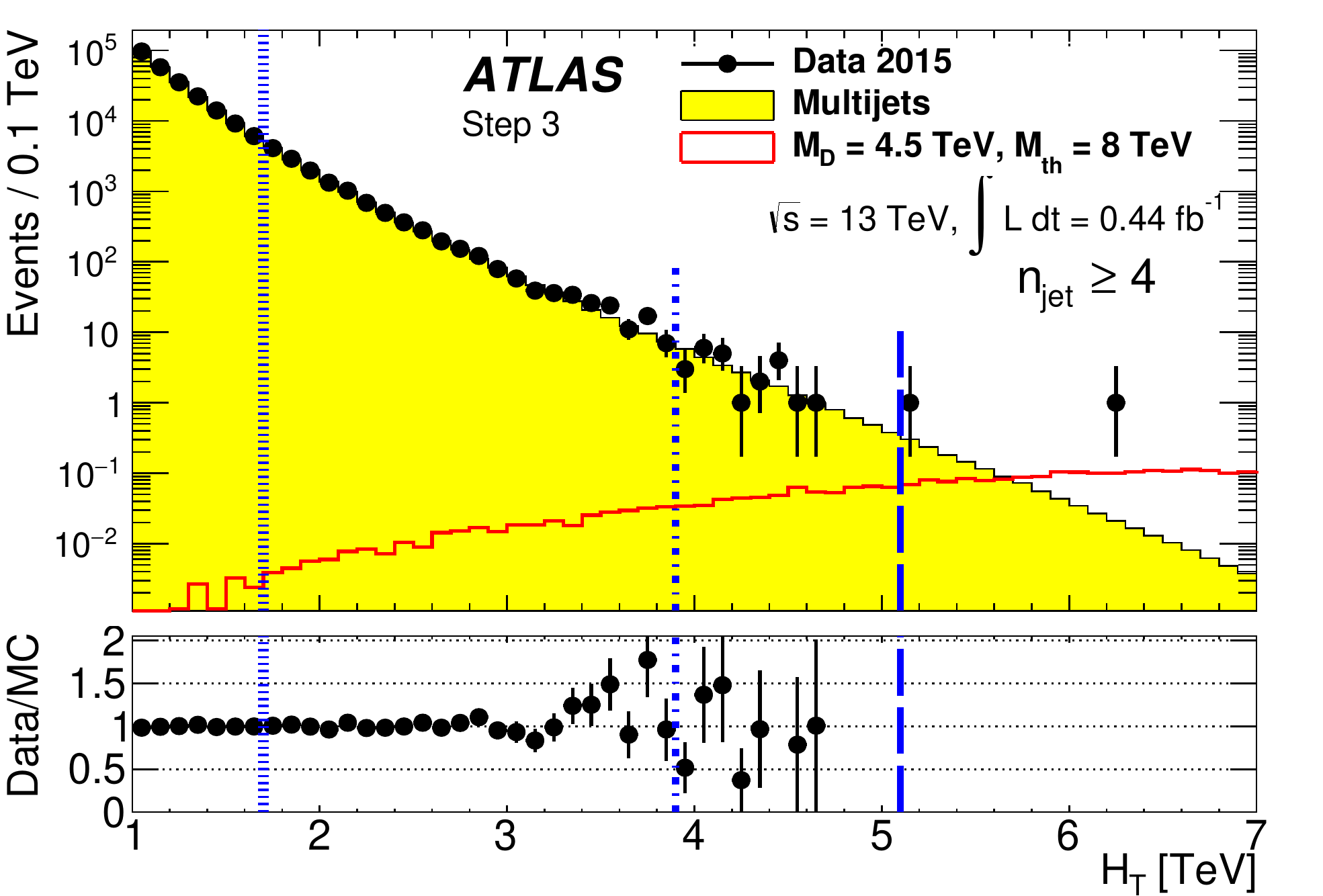}\\
\includegraphics[width=0.48\textwidth]{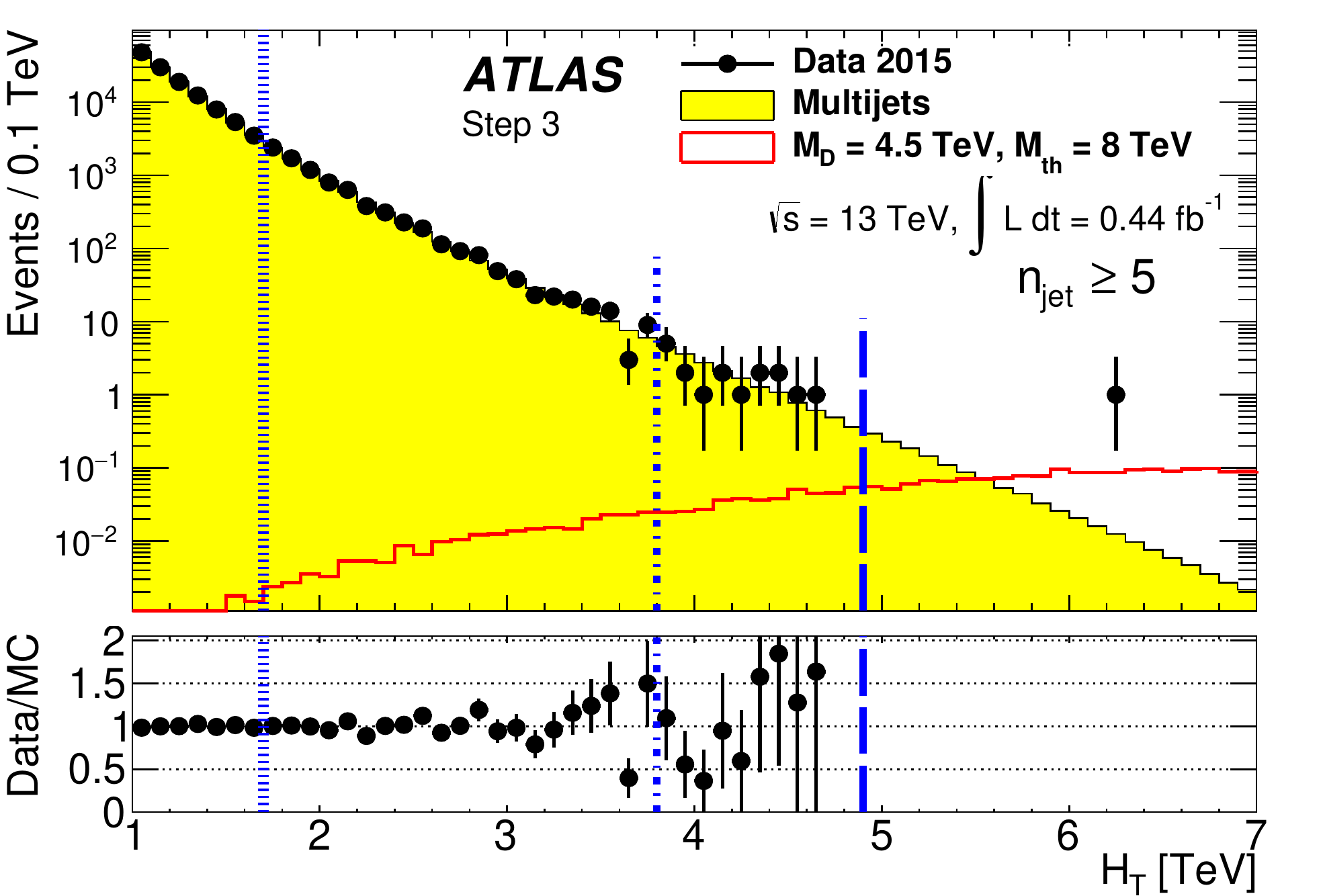}
\includegraphics[width=0.48\textwidth]{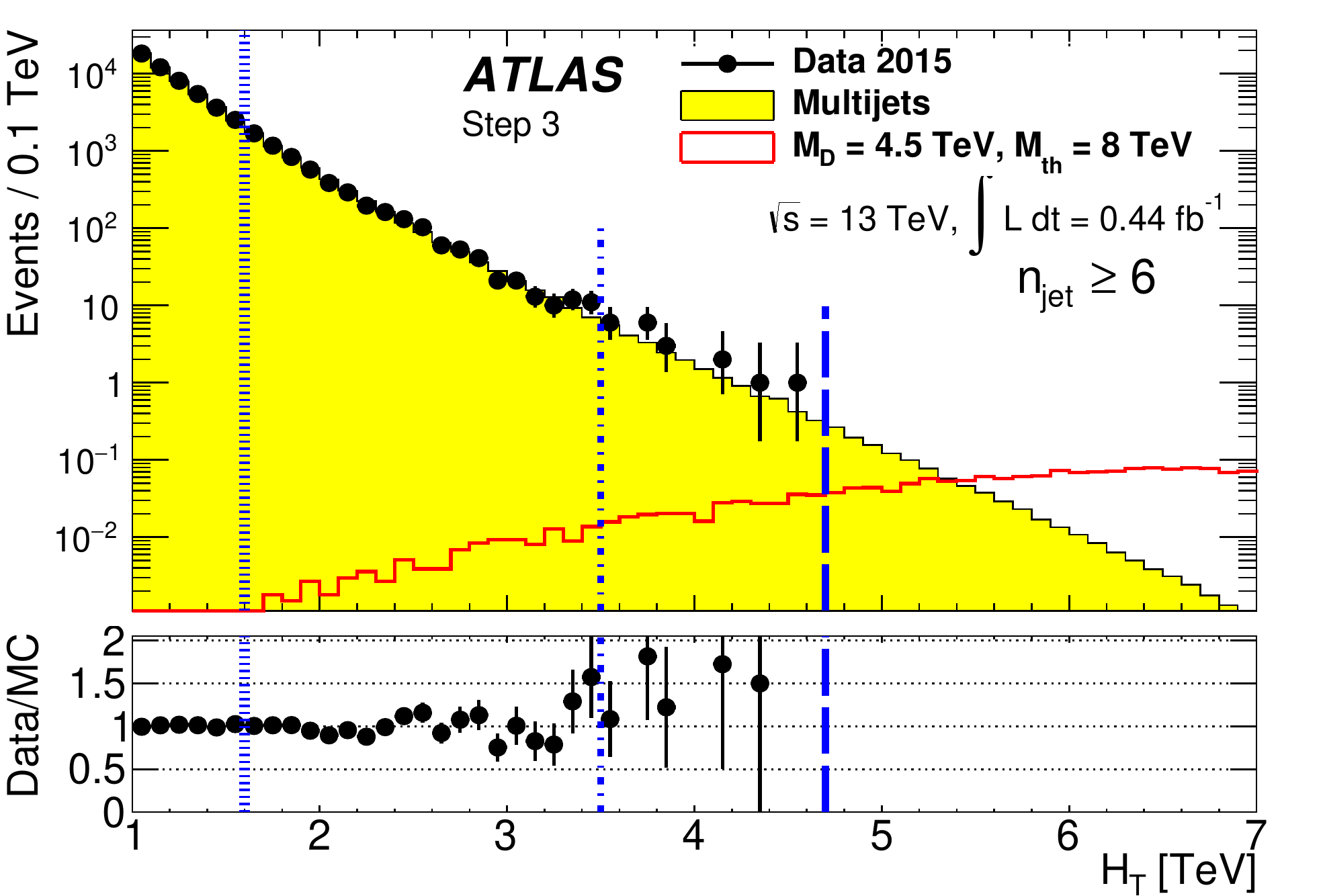}\\
\includegraphics[width=0.48\textwidth]{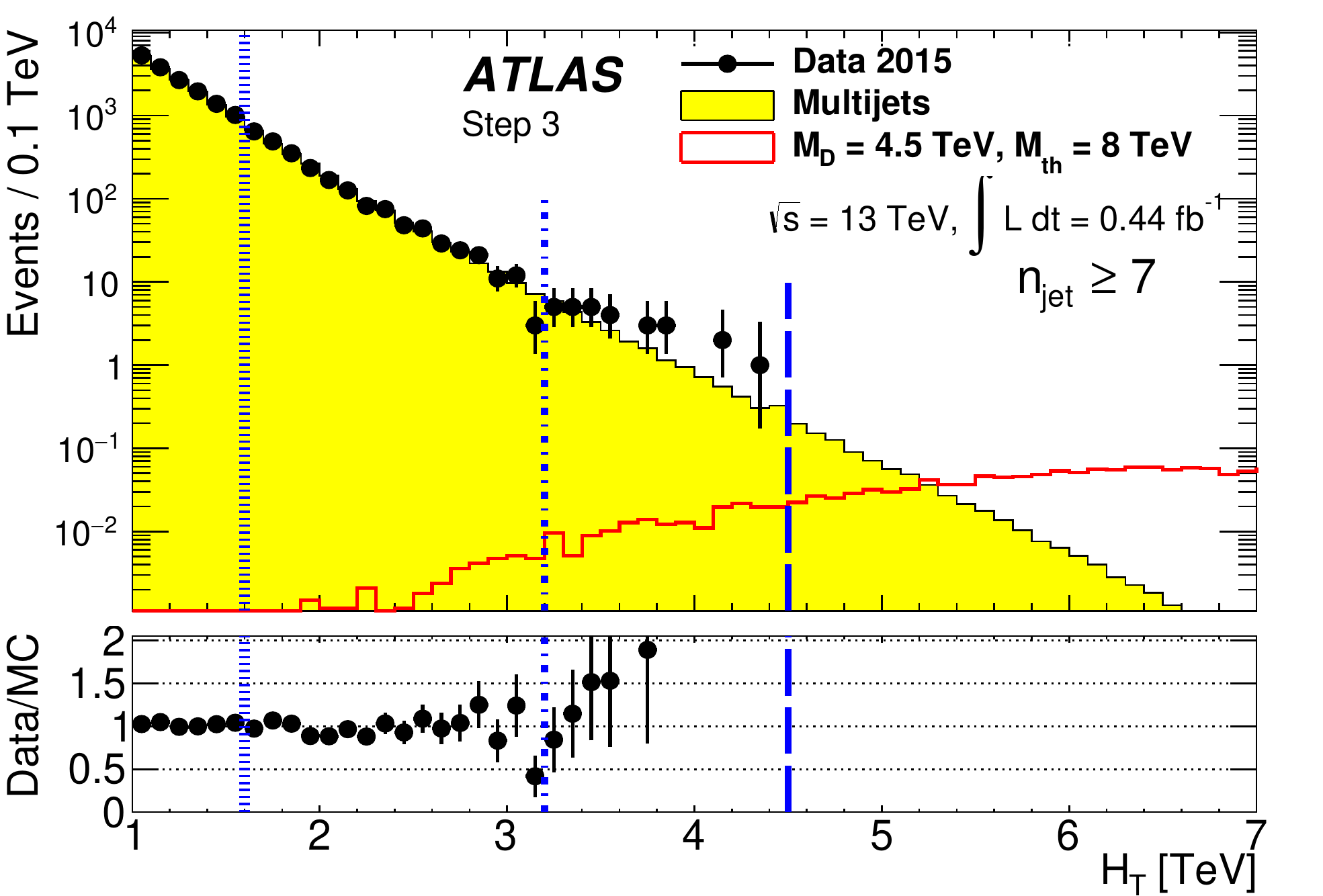}
\includegraphics[width=0.48\textwidth]{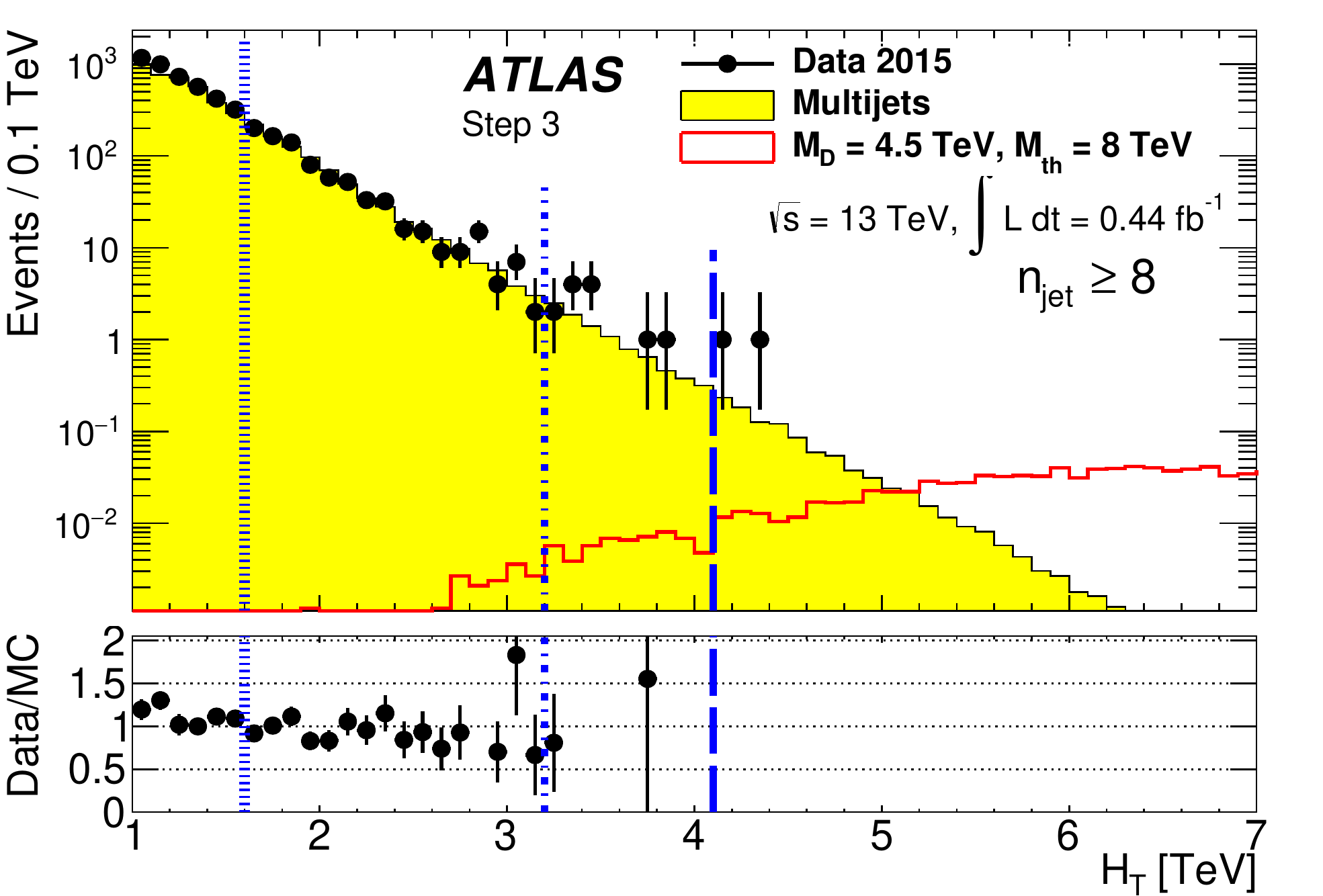}\\
\caption{ Data and MC simulation comparison for \HT{} distributions in different inclusive $\njet$ bins for the \lumic~data sample.
The black hole signal with \MD~=~4.5~\TeV, \Mth~=~8~\TeV~is superimposed with the data and background MC.  The MC simulation is normalized to data in the normalization region.
The vertical dotted line marks the lower boundary of the control region, the vertical dashed-dotted line marks the boundary between control region and validation region, 
and the vertical dashed line marks the boundary between validation region and signal region. These boundaries are determined for each $n_{\mathrm{jet}}$ sample separately.
 }
\label{fig:ht-1000}
\end{center}
\end{figure*}

\begin{figure*}
\begin{center}
\includegraphics[width=0.48\textwidth]{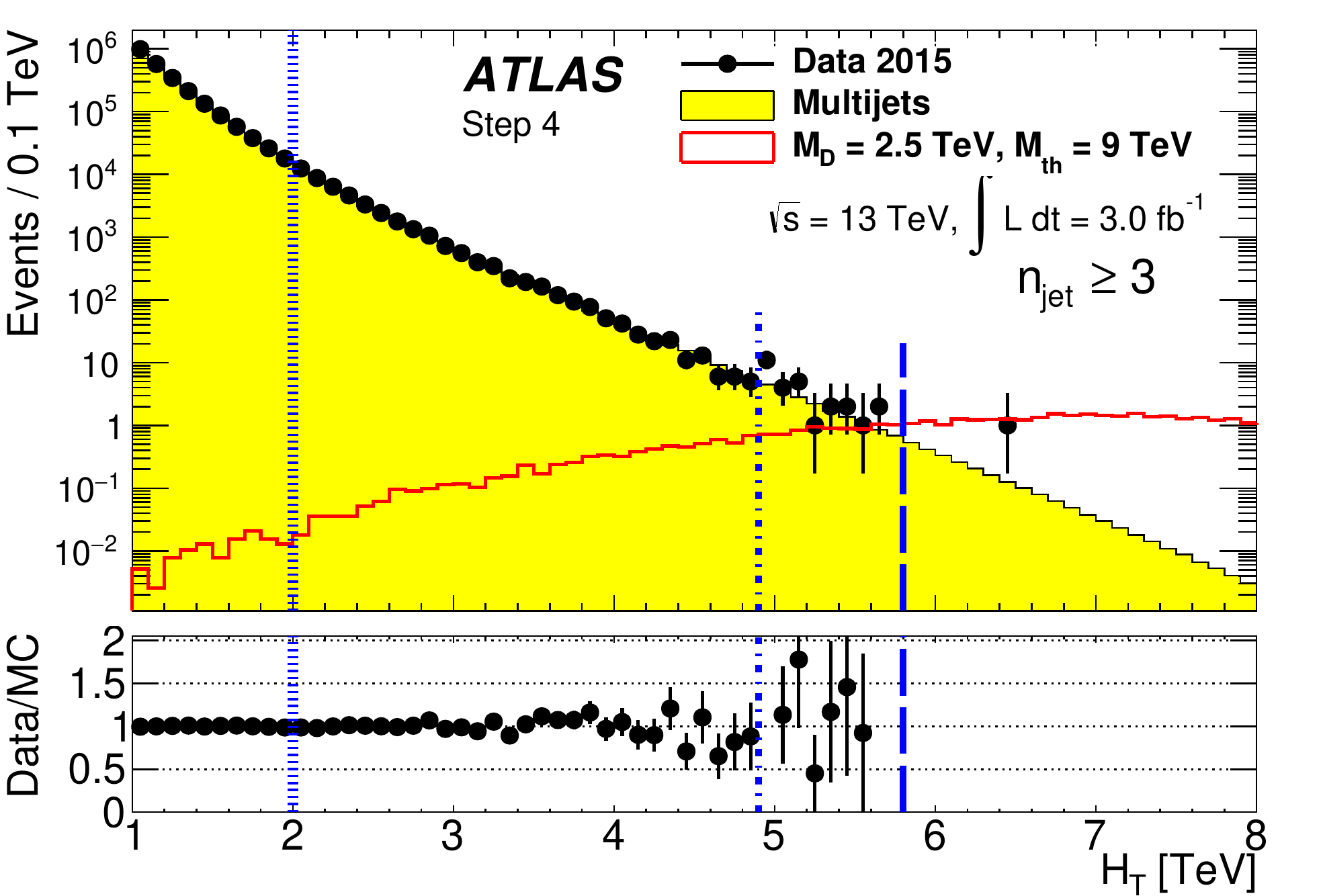}
\includegraphics[width=0.48\textwidth]{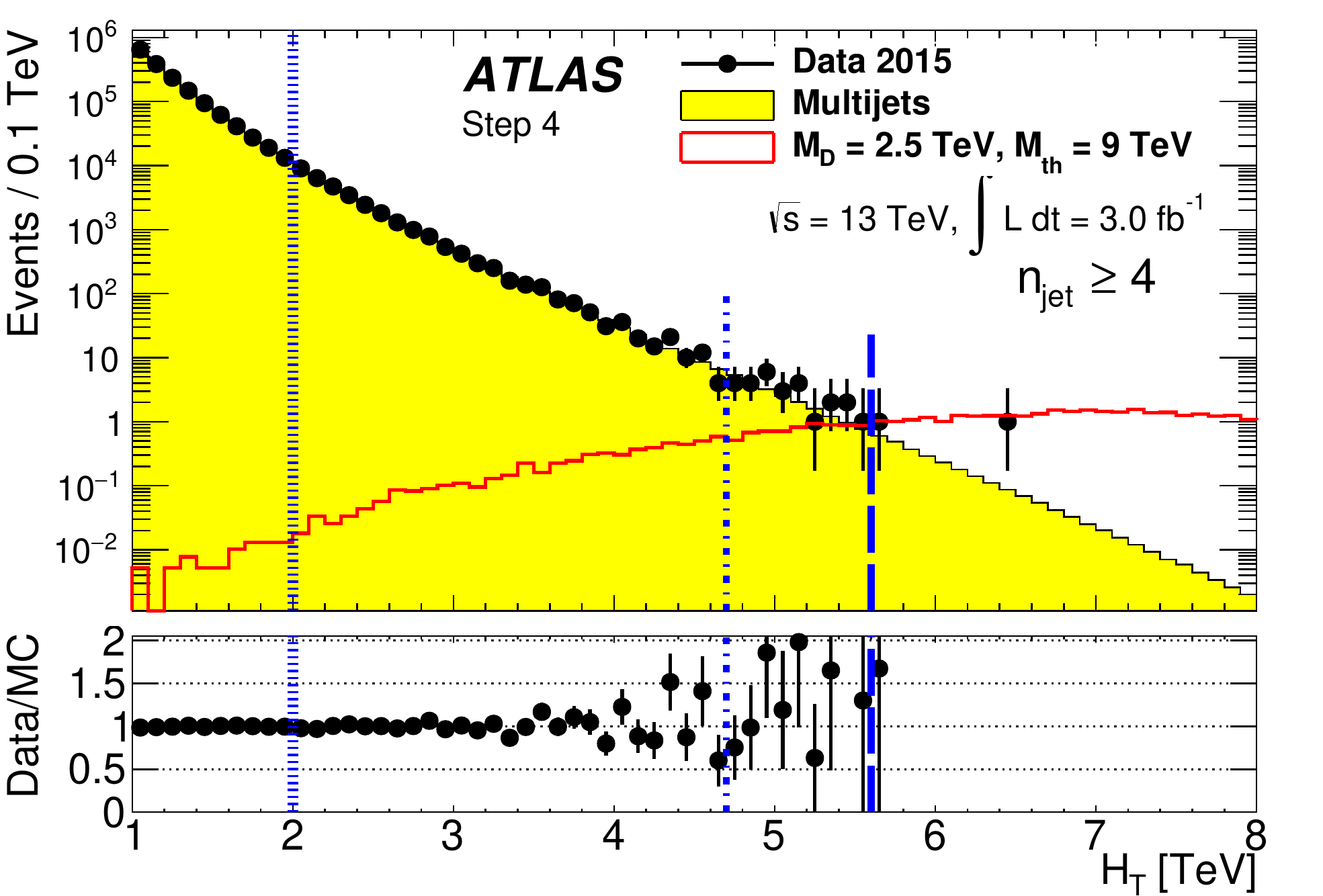}\\
\includegraphics[width=0.48\textwidth]{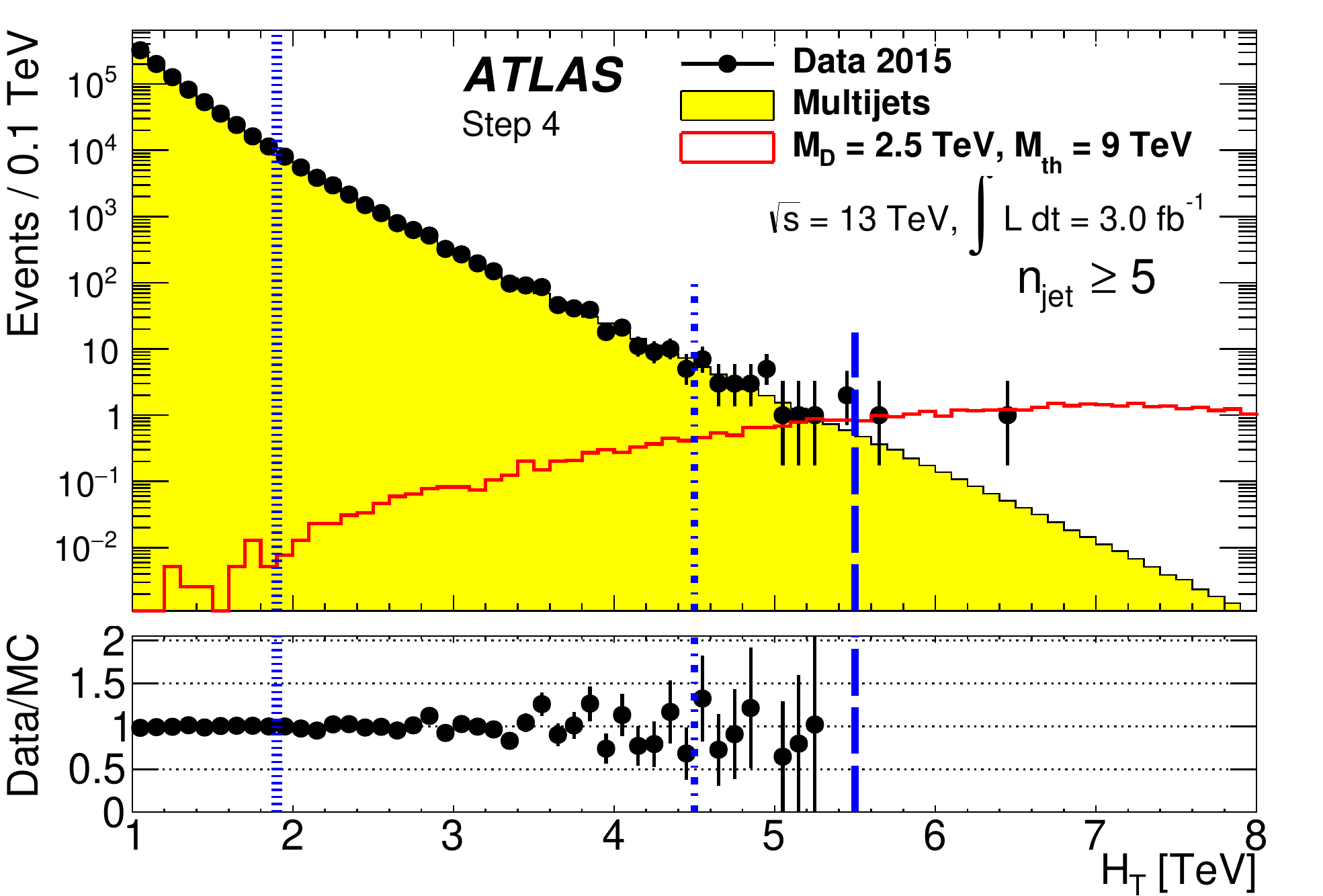}
\includegraphics[width=0.48\textwidth]{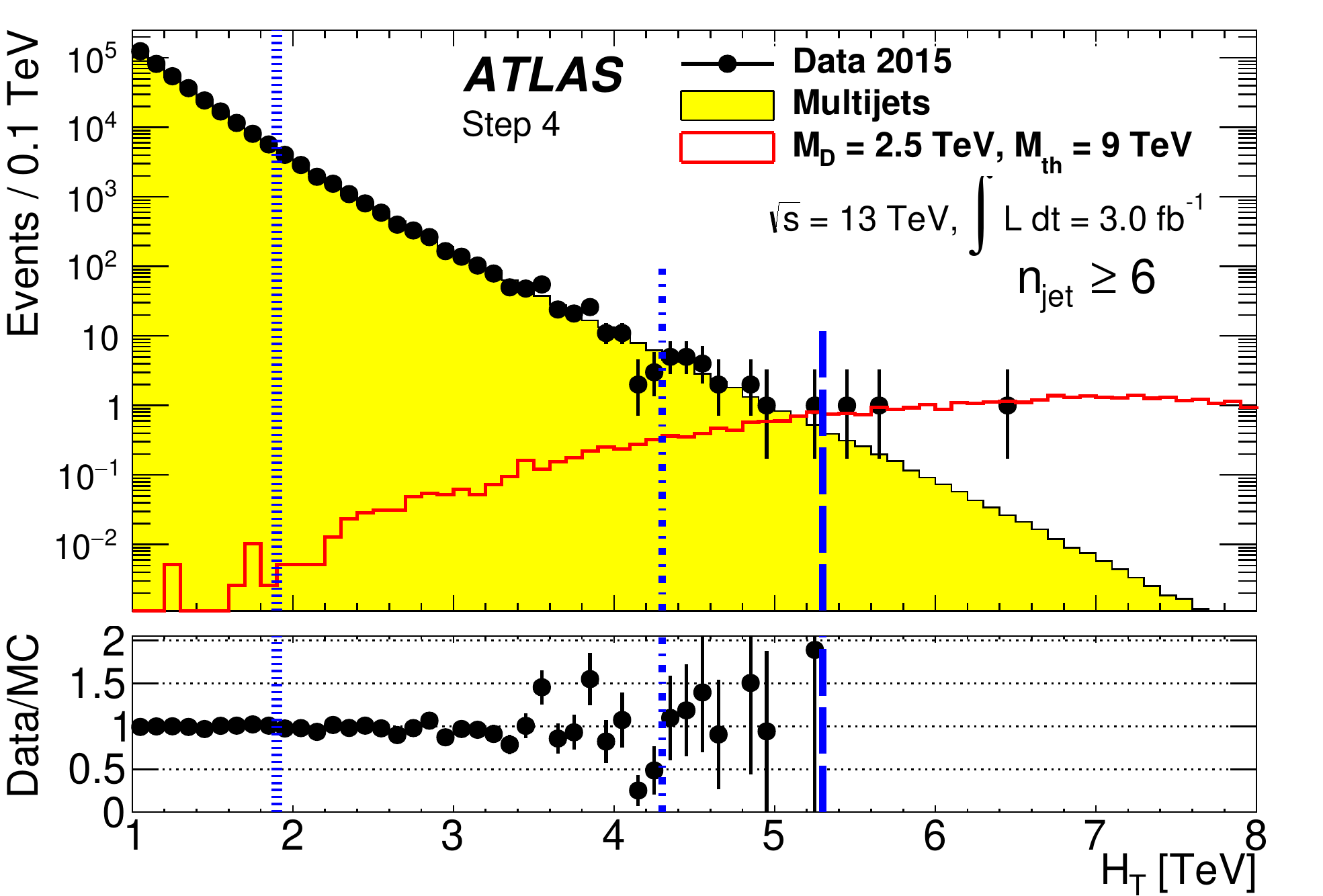}\\
\includegraphics[width=0.48\textwidth]{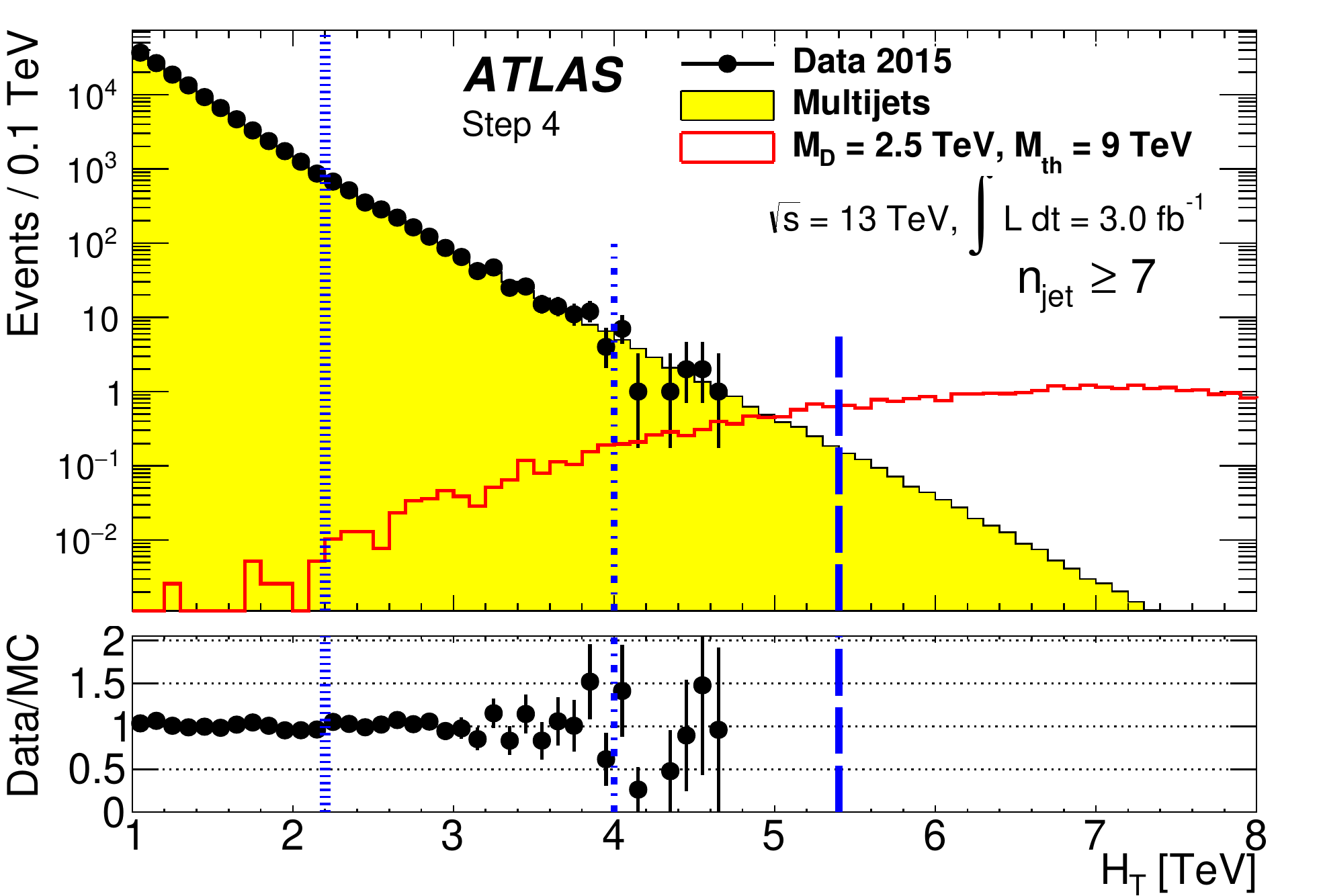}
\includegraphics[width=0.48\textwidth]{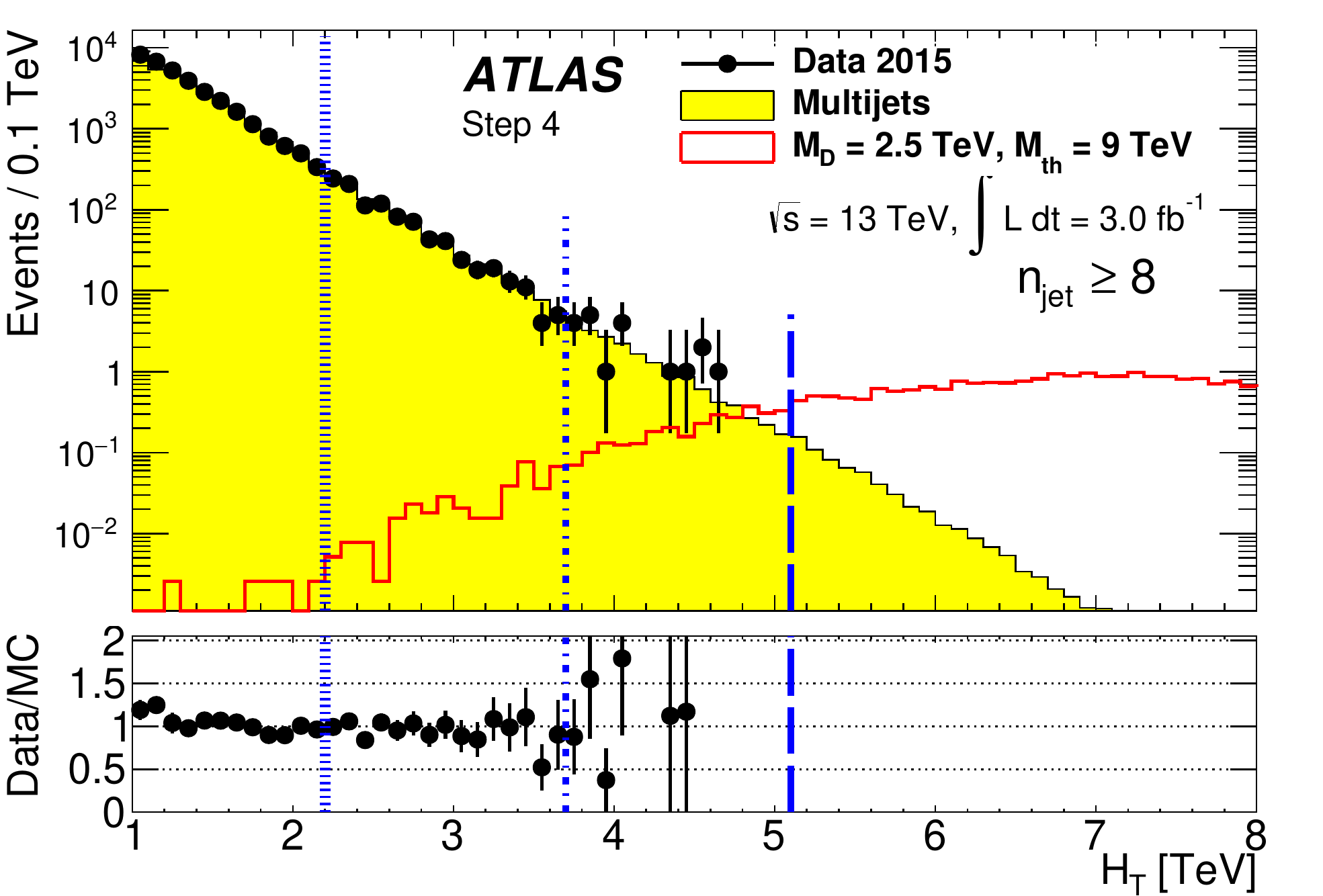}\\

\caption{Data and MC simulation comparison for \HT{} distributions in different inclusive $\njet$ bins for the \lumid~data sample.
The black hole signal with \MD~=~2.5~\TeV, \Mth~=~9.0~\TeV~is superimposed with the data and background MC.  The MC simulation is normalized to data in the normalization region.
The vertical dotted line marks the lower boundary of the control region, the vertical dashed-dotted line marks the boundary between control region and validation region, 
and the vertical dashed line marks the boundary between validation region and signal region. These boundaries are determined for each $n_{\mathrm{jet}}$ sample separately.
 }
\label{fig:ht-10000}
\end{center}
\end{figure*}

As already mentioned, in order to estimate the number of background events in the validation and signal regions, a data-driven method is used.
Data in the control region are fitted to an empirical function which is then used to extrapolate to higher \HT.
The analytic functions considered for this analysis and the allowed ranges of parameters in the fit are summarized in Table~\ref{tab:functions}. 
Function 1 is the baseline function used to fit background for the Run-1 result~\cite{Aad:2015zra}.
Functions 2--10 are the alternative background functions considered or motivated by the Run-1 analysis. 
 These functions are found to fit the \HT~distribution of multijet Monte Carlo samples well and were also used to describe dijet or multijet mass or \HT~distributions in many previous searches~\cite{Alitti:1990kw,Abe:1995jz,Abe:1997hm,Aaltonen:2008dn}.

The 10 functions shown in Table~\ref{tab:functions} are found to fit pseudo-data generated from the simulated multijet sample very well. The difference in fit result between these functions is statistically small and the simulated sample does not have a precision to identify which function is intrinsically better than the rest. Therefore, the choice of the baseline function is not critical. To select a baseline function  in an unbiased manner, the following procedure is applied.
Data corresponding to 1000 pseudo-experiments ~(PEs) drawn from samples of the simulated background 
are used to evaluate the functions and to assess their
ability to obtain a good fit and to correctly predict the event rates in the validation and signal regions. Functions are required to converge in the control region and decrease monotonically with $H_{\mathrm{T}}$ in the signal region for 95\% or more of pseudo-experiments.
Provided these criteria are met,
functions are ranked based on the goodness of their extrapolation in the validation region based on the statistical uncertainty and potential bias of their extrapolation. 
The top-ranked function is selected as the baseline function.
Any function which satisfies these criteria but whose extrapolation does not agree with the data 
in the validation region within 95\% confidence level is rejected and its result is not used to obtain the signal region background uncertainty estimate.

The procedure of ranking and selecting background functions as well as the procedure of determining the control, validation, and signal region boundaries is
repeated for each step used in the bootstrap process and for analyses with different $n_{\mathrm{jet}}$ requirements. 

\begin{table}
\begin{center}
\begin{tabular}{lllclclcl}
\toprule
 &Functional form & $p_1$ & $p_2$ \\
\midrule
1& $f_1(x) = \frac{p_0(1-x)^{p_1}}{x^{p_2}}$ & (0,+$\infty$ )& (0,+$\infty$ )\\ 
\midrule
2& $f_2(x) = p_0(1-x)^{p_1} e^{p_2x^2}$ & (0,+$\infty$ )& (-$\infty$ ,+$\infty$ )\\
\midrule
3& $f_3(x) = p_0(1-x)^{p_1}x^{p_2x}$& (0,+$\infty$ )&(-$\infty$ ,+$\infty$ )\\
\midrule
4& $f_4(x) = p_0 (1-x)^{p_1} x^{p_2\ln x}$& (0,+$\infty$ )&(-$\infty$ ,+$\infty$ )\\
\midrule
5& $f_5(x) = p_0(1-x)^{p_1}(1+x)^{p_2x}$& (0,+$\infty$ )&(0,+$\infty$ )\\
\midrule
6& $f_6(x) = p_0(1-x)^{p_1}(1+x)^{p_2\ln x}$& (0,+$\infty$ )&(0,+$\infty$ )\\
\midrule
7& $f_7(x) = \frac{p_0}{x}(1-x)^{[p_1-p_2\ln x]}$& (0,+$\infty$ )&(0,+$\infty$ )\\
\midrule
8& $f_8(x) = \frac{p_0}{x^2}(1-x)^{[p_1-p_2\ln x]}$& (0,+$\infty$ )&(0,+$\infty$ )\\
\midrule
9& $f_9(x) = \frac{p_0(1-x^{1/3})^{p_1}}{x^{p_2}}$& (0,+$\infty$ )&(0,+$\infty$ )\\
\midrule
10& $f_{10}(x) = p_0 (1-x^{1/3})^{p_1} x^{p_2\ln x}$& (0,+$\infty$ )&(-$\infty$ ,+$\infty$ )\\
\bottomrule
\end{tabular}
\caption{Analytic functions considered in this analysis where $x~=~\HT/\sqrt{s}$. $p_0$ is a normalization constant. 
$p_1$ and $p_2$ are free parameters in a fit, and their allowed floating ranges are shown in the last two columns. 
}
\label{tab:functions}
\end{center}
\end{table}

Figure~\ref{fig:DataCRFit_CRVR_6.6} shows fits to the data in the control region and their extrapolation into the signal 
and validation regions for $\njet\ge 3$ and the data set corresponding to the  first step in the bootstrap. Function 4 is the baseline while functions 1, 9 and 10 pass the goodness of fit and
 monotonicity tests. 
The baseline is used to predict 
rates in the signal region and the others are used to assess systematic uncertainties.
 As will be quantified below, but is already clear from this figure, there is no evidence for a 
discrepancy in the signal and validation regions between the data and the  remaining extrapolations.

Figure~\ref{fig:DataCRFit_CRVR_100} shows the comparison for the \lumib~data set which corresponds to the second step.
Here the functions 1, 3, 4, 5, 6, 9, and 10 in Table~\ref{tab:functions} are qualified for all jet multiplicities with function 10 being the baseline. 
Additionally, function 8 is qualified for $n_{\mathrm{jet}}~\ge~$4 to   $n_{\mathrm{jet}}~\ge~$7, function 7 for $n_{\mathrm{jet}}~\ge~$6  and  $n_{\mathrm{jet}}~\ge~$ 7, and function 2 for  $n_{\mathrm{jet}}~\ge~$ 6.
Figure~\ref{fig:DataCRFit_CRVR_1000} shows the comparison for the \lumic~data set which corresponds to the third step.
Here all the functions  are qualified for all jet multiplicities with 10 being the baseline.
 Finally, Figure~\ref{fig:DataCRFit_CRVR_10000} shows the comparison for the \lumid~data set which corresponds to the fourth step.
Here all functions are qualified for all jet multiplicities less than eight. For $\njet\ge 8$ all functions except 7 and 8 are qualified.
Function 10 is the baseline for all  jet multiplicities except  $n_{\mathrm{jet}}~\ge~$3 where function 5 is the baseline and $n_{\mathrm{jet}}~\ge~$7 where function 4 is   baseline

\begin{figure*}
\begin{center}
\includegraphics[width=1.0\textwidth]{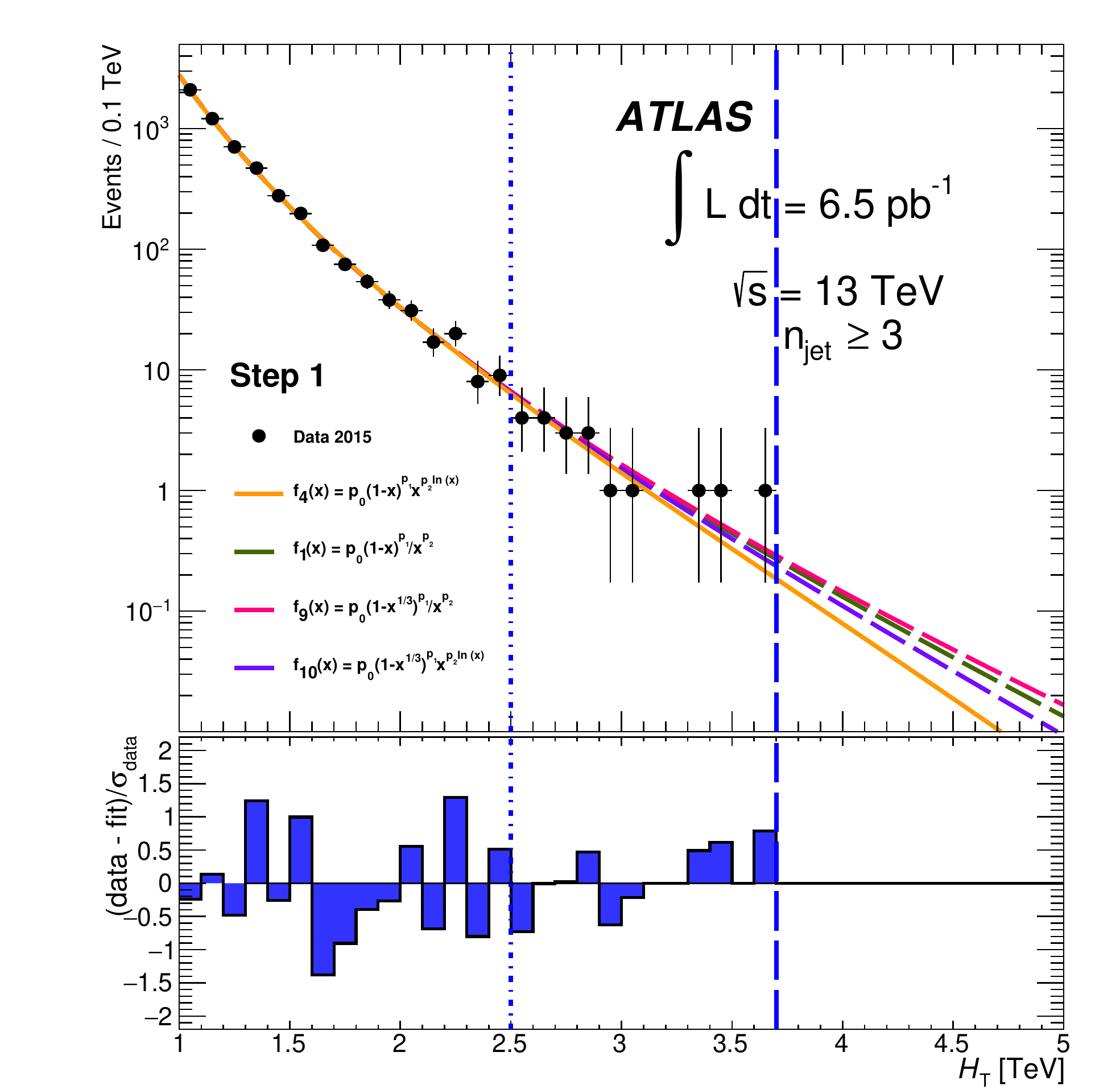}
                                              
\caption{ 
The data in 1.0 \TeV~< $H_{\mathrm{T}}$ < 2.5 \TeV~for  $\njet\ge 3$ are fitted by the baseline function (solid), and three alternative functions (dashed). The fitted functions are
 extrapolated to the validation region and signal region.
The control, validation and 
signal regions are delimited by the vertical lines.
The bottom section of the figure shows the residual significance defined as the ratio of the difference between fit and data over the statistical uncertainty of data, where the fit prediction is taken from the baseline function.
}
\label{fig:DataCRFit_CRVR_6.6} 
\end{center}
\end{figure*}

\begin{figure*}
\begin{center}
\includegraphics[width=1.0\textwidth]{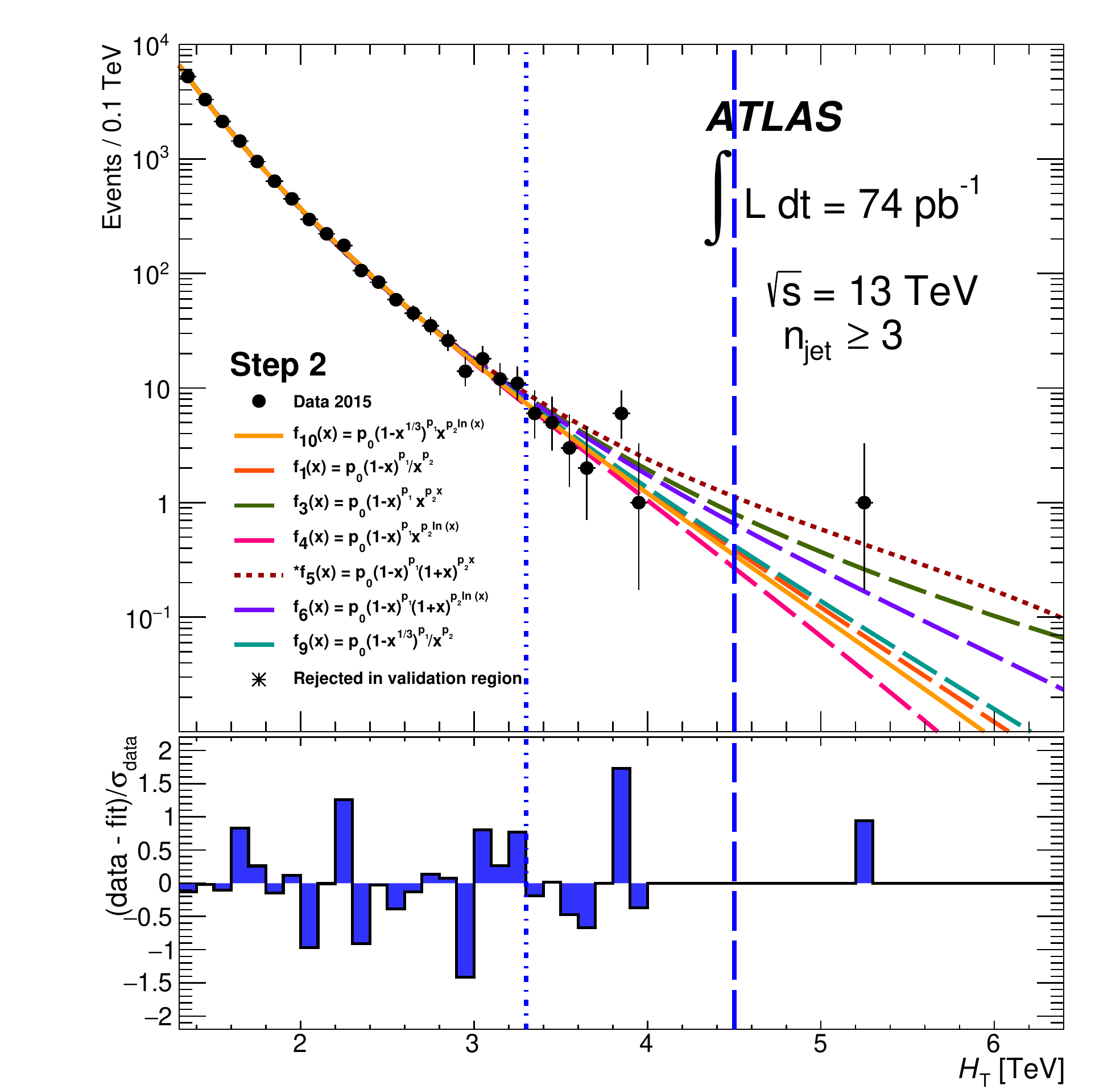}
\caption{ 
The data in 1.2 \TeV~< $H_{\mathrm{T}}$ < 3.3 \TeV~ for  $\njet\ge 3$ are fitted by the baseline function (solid), and six alternative functions (dashed). 
The fitted functions are extrapolated to the validation region and signal region.
 The control, validation and signal regions are delimited by the vertical lines. The function indicated by an asterisk is rejected at 95\% CL by the data in 
the validation region.
The bottom section of the figure shows the residual significance defined as the ratio of the difference between fit and data over the statistical uncertainty of data, where the fit prediction is taken from the baseline function.
}
\label{fig:DataCRFit_CRVR_100} 
\end{center}
\end{figure*}

\begin{figure*}
\begin{center}
\includegraphics[width=1.0\textwidth]{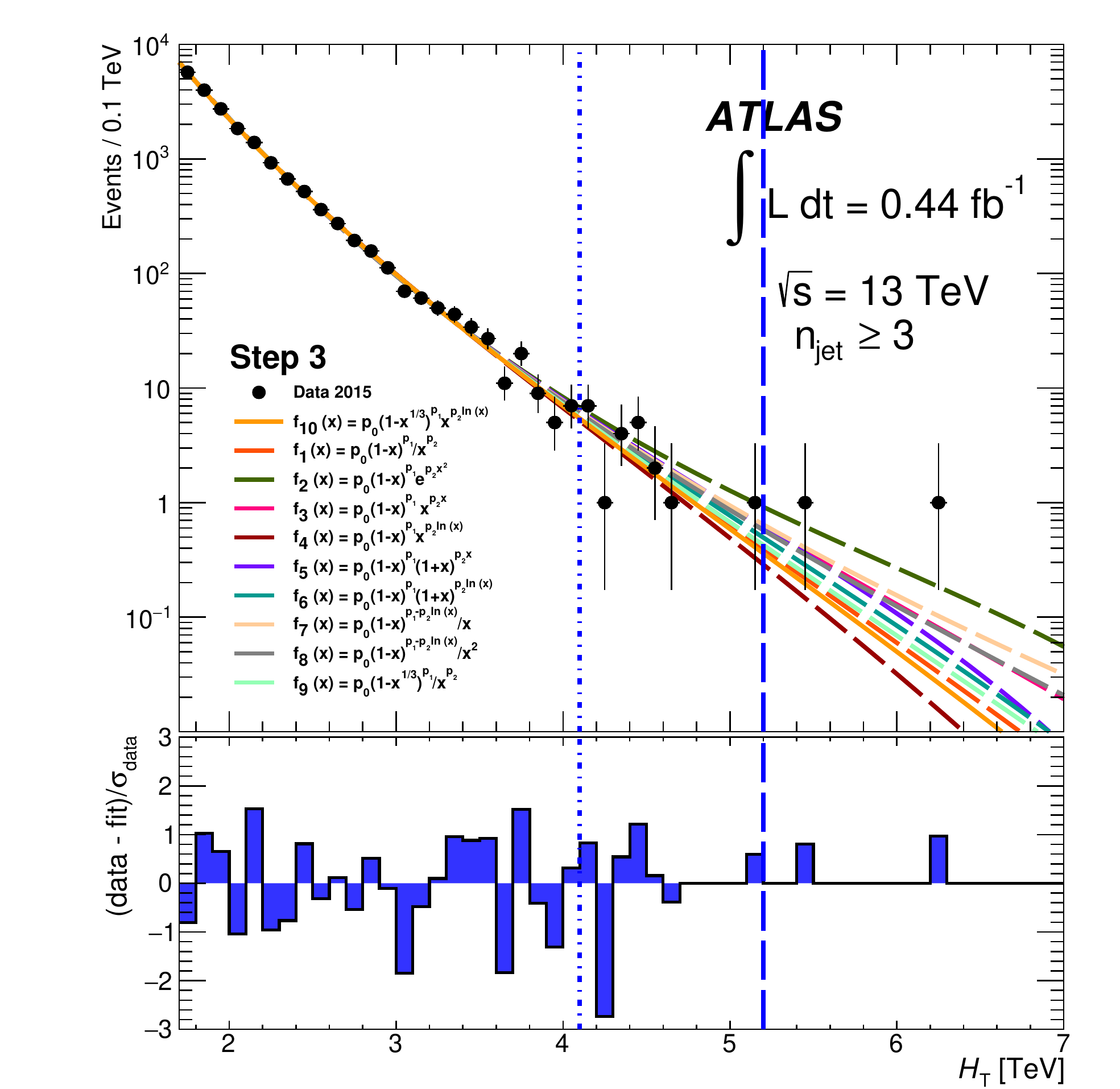}
\caption{ 
The data in 1.7 \TeV~< $H_{\mathrm{T}}$ < 4.1 \TeV~ for  $\njet\ge 3$ are fitted by the baseline function (solid), and nine alternative functions (dashed). 
The fitted functions are extrapolated to the validation region and signal region.
 The control, validation and signal regions are delimited by the vertical lines. 
The bottom section of the figure shows the residual significance defined as the ratio of the difference between fit and data over the statistical uncertainty of data, where the fit prediction is taken from the baseline function.
}
\label{fig:DataCRFit_CRVR_1000} 
\end{center}
\end{figure*}

\begin{figure*}
\begin{center}
\includegraphics[width=1.0\textwidth]{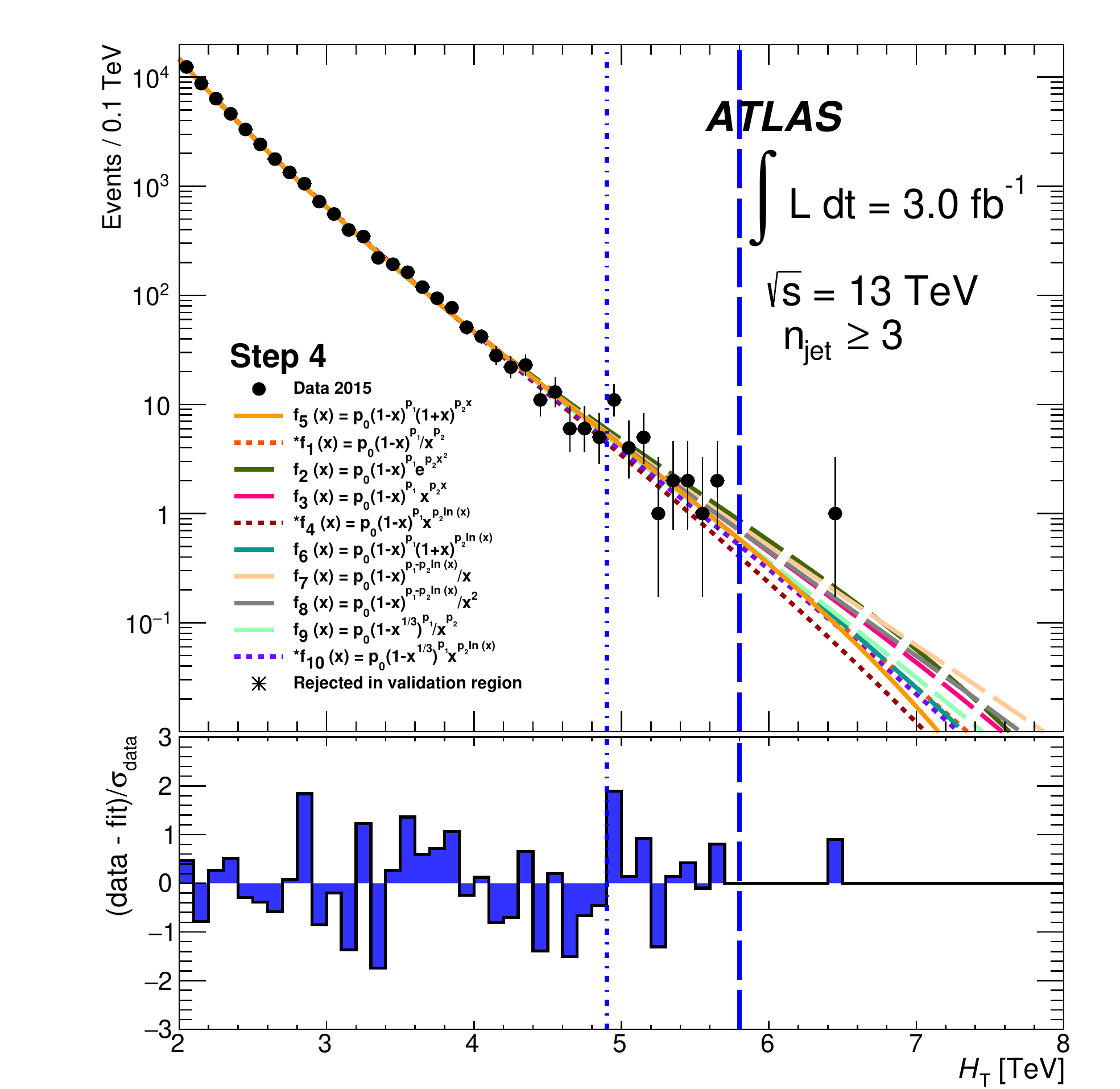}
\caption{ 
The data in 2.0\ TeV < $H_{\mathrm{T}}$ < 4.9 \TeV~ for  $\njet\ge 3$ are fitted by the baseline function (solid), and nine alternative functions (dashed). 
The fitted functions are extrapolated to the validation region and signal region.
 The control, validation and signal regions are delimited by the vertical lines. The three functions indicated by  asterisks are  rejected at 95\% CL by the data in 
the validation region.
The bottom section of the figure shows the residual significance defined as the ratio of the difference between fit and data over the statistical uncertainty of data, where the fit prediction is taken from the baseline function.
}
\label{fig:DataCRFit_CRVR_10000} 
\end{center}
\end{figure*}

\section{Uncertainties}
\label{sec:uncert}
There are two components of uncertainty on the background projections:
a statistical component arising from data fluctuations in the control region
and a systematic component associated with the choice and extrapolation of the empirical fitting functions.
In a pseudo-experiment based approach, 
the statistical component and the extrapolation uncertainty of the baseline fitting function
 are estimated from the width and median value
of the difference between the extrapolations obtained from pseudo-experiments using the baseline function 
and the actual values in the validation and signal regions of the \HT{} distribution.
In a data-driven approach, the maximal difference in the background projection between the baseline function and qualified alternative function 
is used to estimate the uncertainty associated with the choice of fitting function.
The estimated uncertainties are shown in Tables~\ref{tab:events10},  \ref{tab:events100}, \ref{tab:events1000} and \ref{tab:events10000}  where they are indicated by (PE) and (DD) based on the approach used.

In order to convert a limit on the number of events in the signal region to a 
limit on a physics model, simulated signals are needed.
This simulation is used to determine the number of signal events after event 
selection and therefore depends 
on the uncertainties in that determination.
The uncertainty on the expected signal yield includes a luminosity uncertainty of 9\%
 and jet energy scale and resolution uncertainties, which ranges from 1 to 4\% depending on the signal models. 
The latter are critical as they impact the signal 
selection efficiency.

\section{Results}
\label{sec:result}

Tables \ref{tab:events10},  \ref{tab:events100},  \ref{tab:events1000} and  \ref{tab:events10000}
show the predicted number of background events in the validation and signal regions in the data sets corresponding to each step of the analysis.
The first step analysis is shown only for  events with $\njet~\ge~3$: no  useful limit can be obtained for higher multiplicities using this \lumia\ data set as there is 
insufficient data. 

In the case of the second step and   $\njet \ge 3$, function 5
 is excluded at 95\% CL by the observed validation region yield, and the remaining qualified functions (10, 1, 4, 9, 6 and 3) are validated. These are 
shown in Figure~\ref{fig:DataCRFit_CRVR_100}. For the remaining multiplicities all qualified functions are consistent with data in 
the validation region and are used to obtain signal region estimates.

Whether or not any given function can succeed in providing a satisfactory fit depends on the data in the control regions whose boundaries depend on the total luminosity used in that step.
 For the fourth  step and  $\njet \ge 8$, functions 1, 2, 3  4, 5, 6,  9, and 10 are qualified; all the functions are qualified  for the 
remaining jet multiplicities. 
Function 10 is the baseline in all cases except in $\njet \ge 3$ where function 10 as well as functions 1 and 4 are excluded in the validation region. 
The remaining functions (5, 6, 9, 3, 8, 7, 2) are validated, and function 5 becomes the baseline. In other cases, all functions are validated.
These are 
shown in Figure~\ref{fig:DataCRFit_CRVR_10000}. For the remaining multiplicities all validated  functions are consistent with data in 
the validation region and are used to obtain signal region estimates.

 As can be seen from Tables \ref{tab:events10} to \ref{tab:events10000},
the predicted and observed number of events in the validation regions are in
 agreement.  Since there is no excess in the signal region in any given step, there can be no significant signal contributions to the control and validation regions 
for the subsequent steps and limits can be set using the last step where the observation is consistent with the absence of signal.
 The p-values  of a background-only hypothesis  of all the predictions in the signal and validation regions of all the validation functions are larger than 0.1.
The model-dependent 95\% CL limit is shown in Figure~\ref{fig:Limit_MD_Mth_1000} as a
function of $M_{\mathrm{D}}$ and $M_{\mathrm{th}}$ for classical rotating
black holes with $n=2,4,6$ simulated with CHARYBDIS2, using the $n_{\mathrm{jet}} \geq 3$
result for the data set with $3.0$ fb$^{-1}$.   This jet  multiplicity yields 
 the best expected limit for the models under test.
 Limits are shown for classical black holes with $n=2,4$ and 6. 
For the purpose of comparing sensitivity with other LHC searches for strong gravity, 
the interpretation is extended to parameter space where the \Mth~and \MD~are comparable. 
The expected limit significantly exceeds the sensitivity reached by the Run-1 ATLAS search \cite{Aad:2015zra}.
The production of a rotating black hole with n = 6 is excluded, for \Mth~up to 9.0 \TeV--9.7 \TeV, depending on the \MD.
The evolution of the limits with luminosity is shown in Figure~\ref{fig:Limit_with_step1} where a comparison with the Run-1 limit as well as the uncertainty on the final 
expected limit is shown.

\begin{table}
\begin{center}
\begin{tabular}{lclclclclcl}
\toprule
  $\njet\ge$~& VR (obs) & VR (exp) & SR (obs) & SR (exp) \\
\midrule
 3 & 19 & 20.4 $\pm$ 4.4 (PE) $\pm$ 2.6 (DD)   & 0 & 0.65  $\pm$ 0.46 (PE)  $\pm$ 0.64 (DD)  \\
\bottomrule
\end{tabular}
\caption{
The expected and observed number of events in the validation region (VR) and signal region (SR) are shown for $\njet~\ge~3$ in the 6.5 pb$^{-1}$ data set.
The uncertainties on the predicted rates are shown. 
They are obtained from the pseudo-experiment based approach (PE) and the data-driven approach (DD).
}
\label{tab:events10}
\end{center}
\end{table}

\begin{table}
\begin{center}
\begin{tabular}{lclclclclcl}
\toprule
  $\njet\ge$~& VR (obs) & VR (exp) & SR (obs) & SR (exp)\\
\midrule
 3 & 23  & 27.1 $\pm$ 3.7 (PE) $\pm$ 9.6 (DD) & 1 & 1.42 $\pm$ 0.41 (PE)  $^{+4.3}_{-1.42}$ (DD)\\
 4 & 27  & 25.4 $\pm$ 3.2 (PE) $\pm$ 15.5 (DD) & 0 & 1.62 $\pm$ 0.46 (PE) $^{+9.2}_{-1.62}$  (DD)\\
 5 & 21  & 18.9 $\pm$ 2.9 (PE) $\pm$ 9.9 (DD) & 0 & 1.32 $\pm$ 0.48 (PE)  $^{+5.1}_{-1.32}$  (DD)\\
 6 & 18  & 20.7 $\pm$ 3.3 (PE) $\pm$ 10.4 (DD) & 0 & 1.19 $\pm$ 0.48 (PE)  $^{+13.3}_{-1.19}$  (DD)\\
 7 & 29  & 22.2 $\pm$ 3.7 (PE) $\pm$ 7.0 (DD) & 0 & 0.81 $\pm$ 0.36 (PE) $\pm$ 0.60 (DD)\\
\bottomrule
\end{tabular}
\caption{The expected and observed number of events in the validation region (VR) and signal region (SR) are shown for five overlapping inclusive jet multiplicity bins in the  \lumib~data set.
The uncertainties on the predicted rates are shown. 
They are obtained from the pseudo-experiment based approach (PE) and the data-driven approach (DD).
}
\label{tab:events100}
\end{center}
\end{table}

\begin{table}
\begin{center}
\begin{tabular}{lclclclclcl}
\toprule
  $\njet\ge$~& VR (obs) & VR (exp) & SR (obs) & SR (exp)\\
\midrule
 3 & 21  & 20.4 $\pm$ 2.7 (PE) $\pm$ 10.5 (DD) & 2 & 1.46 $\pm$ 0.42 (PE)  $^{+4.37}_{-1.46}$  (DD)\\
 4 & 23 & 29.9 $\pm$ 3.9 (PE) $\pm$ 8.1 (DD) & 2 & 1.95 $\pm$ 0.46 (PE)  $^{+4.06}_{-1.95}$ (DD)\\
 5 & 17 & 21.4 $\pm$ 3.4  (PE) $\pm$ 7.1 (DD) & 1 & 1.56 $\pm$ 0.51 (PE)  $^{+3.47}_{-1.56}$ (DD)\\
 6 & 19 & 28.3  $\pm$ 4.3 (PE) $\pm$ 6.3 (DD) & 0 &   1.44 $\pm$ 0.40 (PE)  $^{+2.13}_{-1.44}$  (DD)\\
 7 & 28 & 24.7 $\pm$ 3.8 (PE) $\pm$ 4.5 (DD) & 0 &   0.96 $\pm$ 0.39 (PE)  $^{+1.74}_{-0.96}$  (DD)\\
8  & 25 & 31.8 $\pm$ 4.7 (PE) $\pm$ 1.4 (DD) & 2 &   2.86 $\pm$ 0.40 (PE) $\pm$ 0.70 (DD)\\
\bottomrule
\end{tabular}
\caption{The expected and observed number of events in the validation region (VR) and signal region (SR) are shown for six overlapping inclusive jet multiplicity bins in the  \lumic~data set.
The uncertainties on the predicted rates are shown. 
They are obtained from the pseudo-experiment based approach (PE) and the data-driven approach (DD).
}
\label{tab:events1000}
\end{center}
\end{table}

\begin{table}
\begin{center}
\begin{tabular}{lclclclclcl}
\toprule
  $\njet\ge$~& VR (obs) & VR (exp) & SR (obs) & SR (exp)\\
\midrule
 3 & 28 & 19.5 $\pm$ 3.6 (PE) $\pm$ 4.1 (DD) & 1 & 2.10 $\pm$ 0.51 (PE) $\pm$ 1.78 (DD)\\
 4 & 27 & 20.8 $\pm$ 2.3 (PE) $\pm$ 6.4 (DD) & 2 & 2.36 $\pm$ 0.52 (PE) $\pm$ 2.12 (DD)\\
 5 & 26 & 22.3 $\pm$ 2.6  (PE) $\pm$ 6.8 (DD) & 2 & 1.95 $\pm$ 0.45 (PE) $^{+2.10}_{-1.95}$ (DD)\\
 6 & 20 & 20.3  $\pm$ 2.9 (PE) $\pm$ 5.4 (DD) & 3 &   1.82 $\pm$ 0.49 (PE) $^{+1.91}_{-1.82}$ (DD)\\
 7 & 14 & 20.7 $\pm$ 4.1 (PE) $\pm$ 1.7 (DD) & 0 & 0.53 $\pm$ 0.36 (PE) $\pm$ 0.22 (DD)\\
 8 & 19 & 18.2 $\pm$ 4.9 (PE) $\pm$ 3.5 (DD) & 0 &   0.43 $\pm$ 0.36 (PE) $\pm$ 0.26 (DD)\\
\bottomrule
\end{tabular}
\caption{The expected and observed number of events in the validation region (VR) and signal region (SR) are shown for six overlapping inclusive jet 
multiplicity bins in the  \lumid~data set.
The uncertainties on the predicted rates are shown. 
They are obtained from the pseudo-experiment based approach (PE) and the data-driven approach (DD).
}
\label{tab:events10000}
\end{center}
\end{table}

\begin{figure*}
\begin{center}
\includegraphics[width=0.88\textwidth]{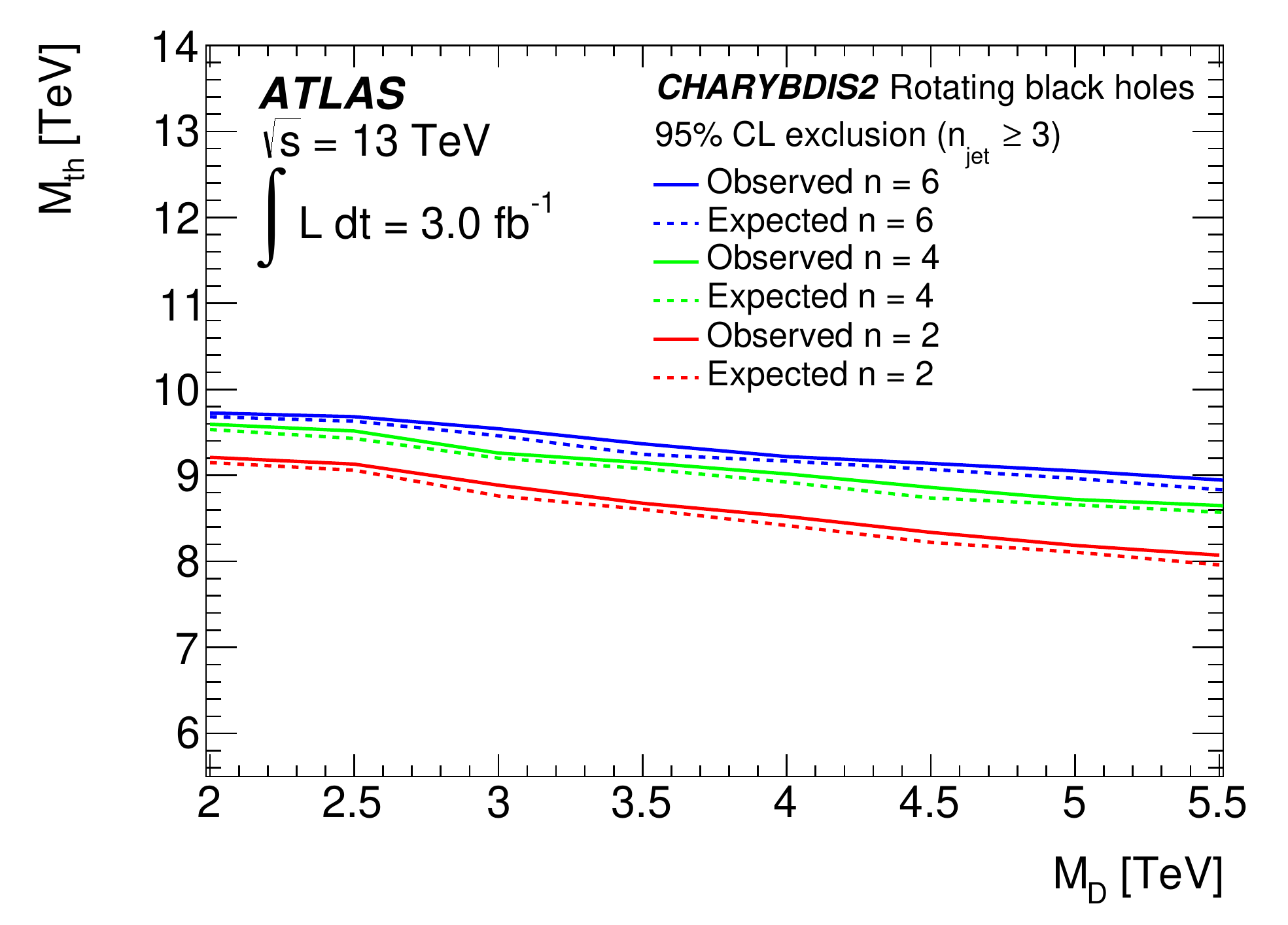}
\caption{The observed and expected 95\% CL exclusion limits on rotating black holes
with different numbers of extra dimensions ($n=2,4,6$) in the $M_{\mathrm{D}}-M_{\mathrm{th}}$ grid.
The results are based on the analysis of \lumid~of integrated luminosity. The region below the lines is excluded.
}
\label{fig:Limit_MD_Mth_1000}
\end{center}
\end{figure*}

An interpretation in terms of the string ball model with six extra dimensions is shown in Figure~\ref{fig:stringballlimit}. Limits are shown as a function of \Mth~and
\gs~for constant \Ms~and as a function of \Mth~and \Ms~for fixed \gs.

\begin{table}
\begin{center}
\begin{tabular}{lclclclclcl}
\toprule
  $\njet\ge$ & $\HT>H_{\mathrm{T}}^{\mathrm{min}}$ (\TeV) & Expected limit (fb) & Observed limit (fb)\\
\midrule
 3 &  5.8 & 1.63$^{+0.70}_{-0.57}$ & 1.33 \\
 4 & 5.6 & 1.77$^{+0.70}_{-0.57}$  & 1.77 \\
 5 & 5.5  & 1.56$^{+0.73}_{-0.50}$  & 1.75 \\
 6 & 5.3  & 1.52$^{+0.69}_{-0.50}$  & 2.15 \\
 7 & 5.4  & 1.02$^{+0.36}_{-0.0}$  &  1.02  \\
 8 & 5.1 & 1.01$^{+0.29}_{-0.0}$  &  1.01 \\

\bottomrule
\end{tabular}
\caption{The expected and observed limits on the inclusive cross section in femtobarns for  production of events as a function of $\njet$  and the minimum value of $\HT$ . 
The limits are derived from results of the \lumid~analysis so $H_{\mathrm{T}}^{\mathrm{min}}$ corresponds to the value of $S$ for the last analysis step.}
\label{tab:sigmalim}
\end{center}
\end{table}

The limits can be re-expressed in terms of a limit on the cross section to produce new physics with a minimum $\HT$~requirement ($H_{\mathrm{T}}^{\mathrm{min}}$) as a function  
of $\njet$ with the kinematic restriction that each jet must satisfy  $\pT > 50\GeV$ and $|\eta|<2.8$ and that  at least one jet must have
 $\pT > 200\GeV$. In order to do this the efficiency for detecting events satisfying this kinematic requirement must be known.
This efficiency is model-dependent. 
A conservative estimate was obtained by taking the minimal efficiency from signal models whose predicted rates lie within $\pm$ 10\% of the observed limits.
The minimum efficiency is found to be 0.98. 
The resulting limit on the cross section is shown in Table~\ref{tab:sigmalim} 
which shows the expected limits together with their uncertainties, and the observed limits.

\begin{figure*}
\begin{center}
\includegraphics[width=0.88\textwidth]{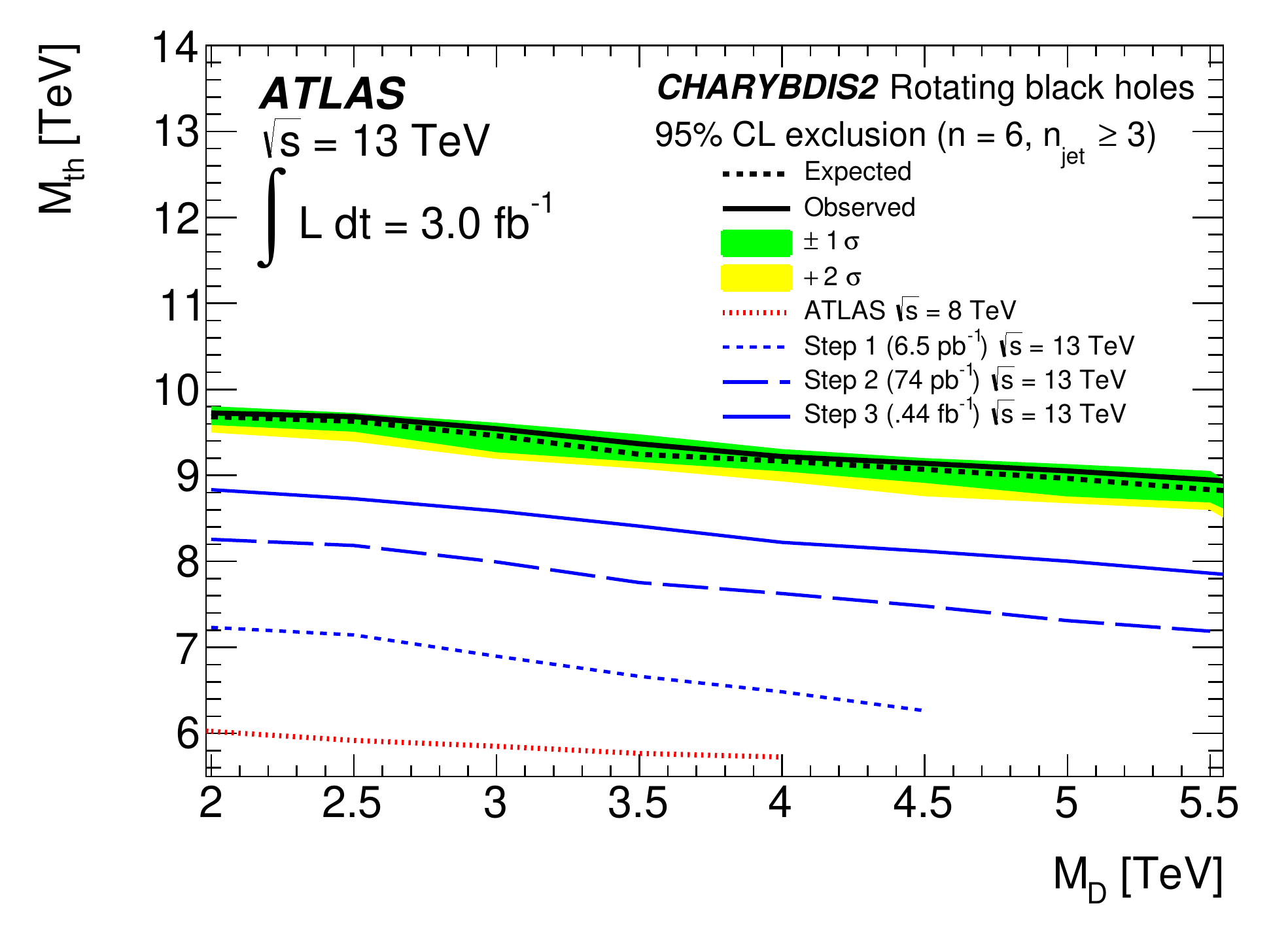}
\caption{ "The observed and expected limits on rotating black holes
with $n=6$ in the $M_{\mathrm{D}}-M_{\mathrm{th}}$ grid,  from the analysis with 
an integrated luminosity of \lumid. The 95\% CL expected limit is shown as the black dashed line, 
and limits corresponding to the  $\pm~1~\sigma$ and $+~2~\sigma$ variations of the background 
expectation are shown as the green and yellow bands, respectively. The 95\% CL observed limit is shown as the black solid line. 
 The $-2~\sigma$ band is not shown as it almost completely overlaps with the $-1~\sigma$ band.
The blue dashed lines corresponds to the observed limits from the first, second  and third step analyses.
The red dotted line corresponds to the limit from Run-1 ATLAS multijet search~\cite{Aad:2015zra}. }
\label{fig:Limit_with_step1}
\end{center}
\end{figure*}

\begin{figure*}
\begin{center}
\includegraphics[width=0.48\textwidth]{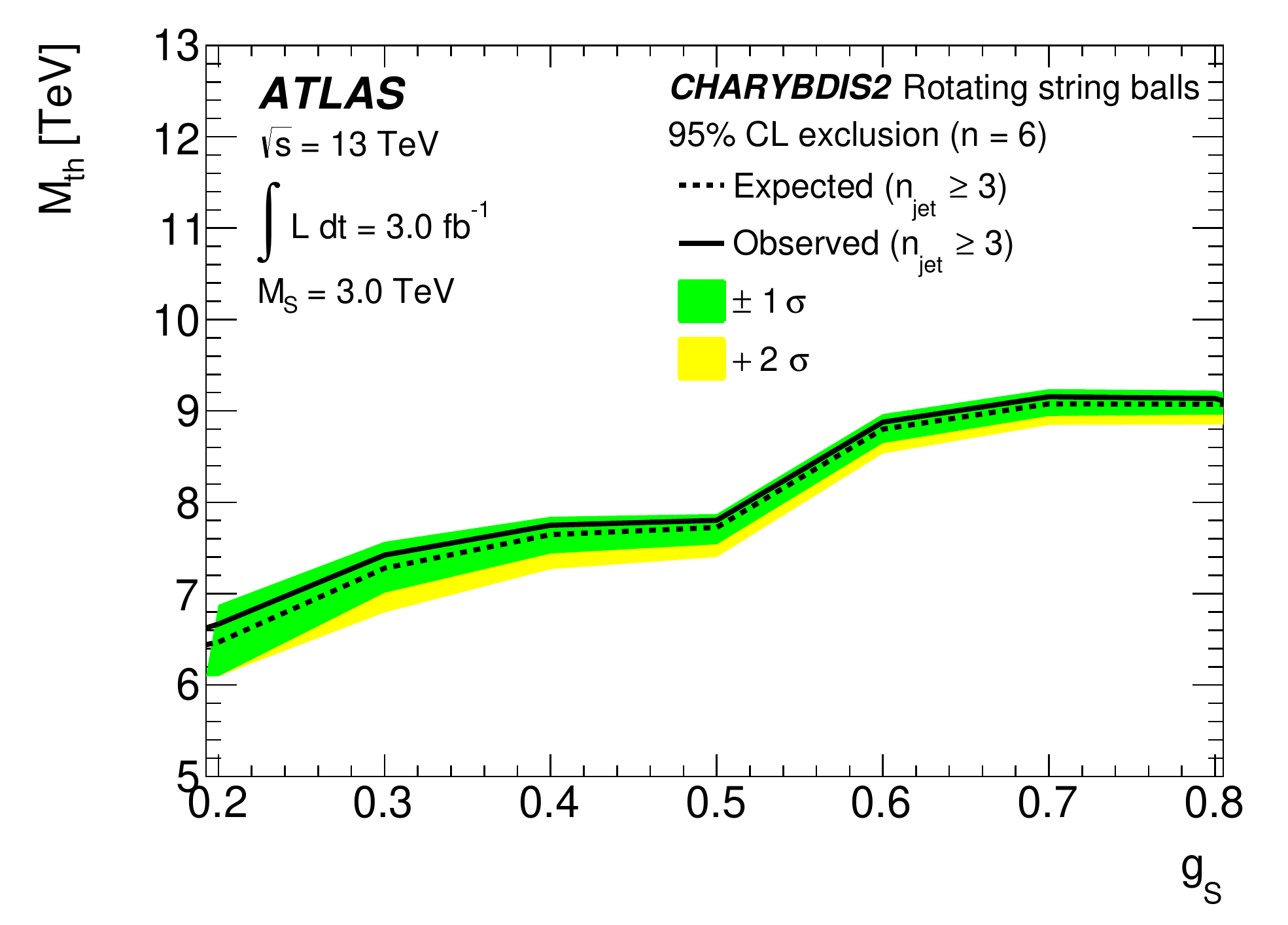} 
\includegraphics[width=0.48\textwidth]{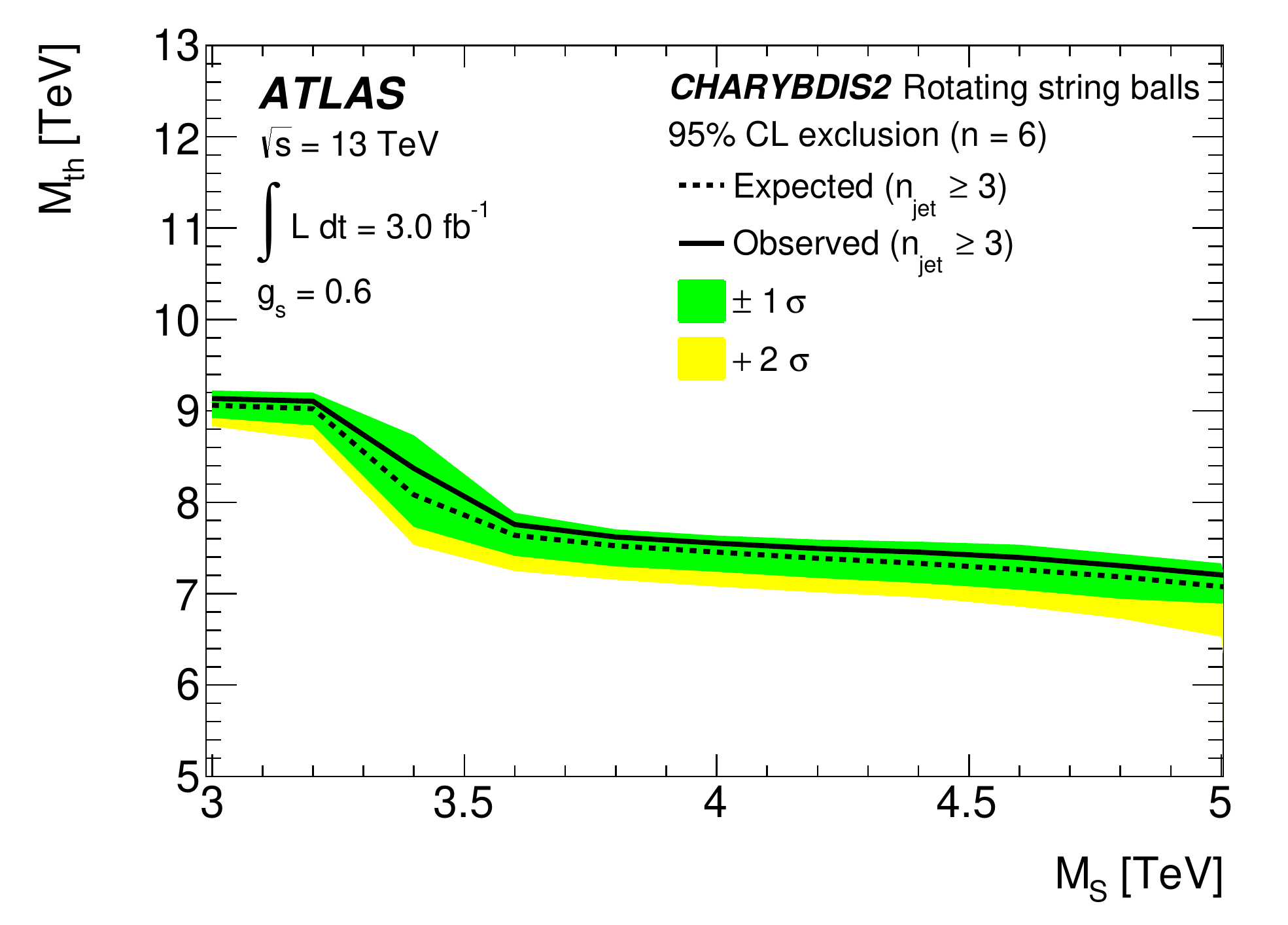}
\caption{ 
The expected and observed limits on the string ball model with $n=6$, from the analysis with 
an integrated luminosity of \lumid. The left plot shows  the 95\% CL limit as a function of \gs~and \Mth~(solid line).
 The  dashed line shows the  expected limit; the  limits corresponding to the  $\pm~1~\sigma$ and $+~2~\sigma$ variations of the background 
expectation are shown as the green and yellow bands, respectively.
 The $-2~\sigma$ band is not shown as it almost completely overlaps with the $-1~\sigma$ band.
The right plot shows the limits as a function of \Mth~and \Ms~for \gs~=~0.6.  }
\label{fig:stringballlimit}
\end{center}
\end{figure*}

 \FloatBarrier

\section{Conclusion}
\label{sec:conclusion}

A search for signals of strong gravity in multijet final states was\ performed
using \lumi~of proton--proton data taken at 13 \TeV~from the Large Hadron Collider using the ATLAS detector. Distributions of events as a function of the scalar sum
of the transverse momenta of jets were examined.  No evidence for
 deviations from Standard Model  expectations at large $\HT$  has been seen. In the \charybdis~1.0.4 model exclusions are shown 
as a function of \MD\  and  \Mth. 
The production of a rotating black hole with $n$ = 6 is excluded, for \Mth\ up to 9.0 \TeV--9.7 \TeV, depending on the \MD. Limits on parameters in the string-ball model are also set.
 These extend significantly the limits from the 8 \TeV~LHC analyses.

\section*{Acknowledgments}


We thank CERN for the very successful operation of the LHC, as well as the
support staff from our institutions without whom ATLAS could not be
operated efficiently.

We acknowledge the support of ANPCyT, Argentina; YerPhI, Armenia; ARC, Australia; BMWFW and FWF, Austria; ANAS, Azerbaijan; SSTC, Belarus; CNPq and FAPESP, Brazil; NSERC, NRC and CFI, Canada; CERN; CONICYT, Chile; CAS, MOST and NSFC, China; COLCIENCIAS, Colombia; MSMT CR, MPO CR and VSC CR, Czech Republic; DNRF, DNSRC and Lundbeck Foundation, Denmark; IN2P3-CNRS, CEA-DSM/IRFU, France; GNSF, Georgia; BMBF, HGF, and MPG, Germany; GSRT, Greece; RGC, Hong Kong SAR, China; ISF, I-CORE and Benoziyo Center, Israel; INFN, Italy; MEXT and JSPS, Japan; CNRST, Morocco; FOM and NWO, Netherlands; RCN, Norway; MNiSW and NCN, Poland; FCT, Portugal; MNE/IFA, Romania; MES of Russia and NRC KI, Russian Federation; JINR; MESTD, Serbia; MSSR, Slovakia; ARRS and MIZ\v{S}, Slovenia; DST/NRF, South Africa; MINECO, Spain; SRC and Wallenberg Foundation, Sweden; SERI, SNSF and Cantons of Bern and Geneva, Switzerland; MOST, Taiwan; TAEK, Turkey; STFC, United Kingdom; DOE and NSF, United States of America. In addition, individual groups and members have received support from BCKDF, the Canada Council, CANARIE, CRC, Compute Canada, FQRNT, and the Ontario Innovation Trust, Canada; EPLANET, ERC, FP7, Horizon 2020 and Marie Skłodowska-Curie Actions, European Union; Investissements d'Avenir Labex and Idex, ANR, Region Auvergne and Fondation Partager le Savoir, France; DFG and AvH Foundation, Germany; Herakleitos, Thales and Aristeia programmes co-financed by EU-ESF and the Greek NSRF; BSF, GIF and Minerva, Israel; BRF, Norway; the Royal Society and Leverhulme Trust, United Kingdom.

The crucial computing support from all WLCG partners is acknowledged
gratefully, in particular from CERN and the ATLAS Tier-1 facilities at
TRIUMF (Canada), NDGF (Denmark, Norway, Sweden), CC-IN2P3 (France),
KIT/GridKA (Germany), INFN-CNAF (Italy), NL-T1 (Netherlands), PIC (Spain),
ASGC (Taiwan), RAL (UK) and BNL (USA) and in the Tier-2 facilities
worldwide.

\clearpage

\printbibliography

\clearpage
\vfill 
\eject

\clearpage
\begin{flushleft}
{\Large The ATLAS Collaboration}

\bigskip

G.~Aad$^\textrm{\scriptsize 85}$,
B.~Abbott$^\textrm{\scriptsize 112}$,
J.~Abdallah$^\textrm{\scriptsize 150}$,
O.~Abdinov$^\textrm{\scriptsize 11}$,
B.~Abeloos$^\textrm{\scriptsize 116}$,
R.~Aben$^\textrm{\scriptsize 106}$,
M.~Abolins$^\textrm{\scriptsize 90}$,
O.S.~AbouZeid$^\textrm{\scriptsize 157}$,
H.~Abramowicz$^\textrm{\scriptsize 152}$,
H.~Abreu$^\textrm{\scriptsize 151}$
% ...
\bigskip
\\
% ...
$^{11}$ Institute of Physics, Azerbaijan Academy of Sciences, Baku, Azerbaijan\\
% ...
$^{85}$ CPPM, Aix-Marseille Universit{\'e} and CNRS/IN2P3, Marseille, France\\
% ...
$^{90}$ Department of Physics and Astronomy, Michigan State University, East Lansing MI, United States of America\\
% ...
$^{106}$ Nikhef National Institute for Subatomic Physics and University of Amsterdam, Amsterdam, Netherlands\\
% ...
$^{112}$ Homer L. Dodge Department of Physics and Astronomy, University of Oklahoma, Norman OK, United States of America\\
% ...
$^{116}$ LAL, Univ. Paris-Sud, CNRS/IN2P3, Universit{\'e} Paris-Saclay, Orsay, France\\
% ...
$^{150}$ Institute of Physics, Academia Sinica, Taipei, Taiwan\\
$^{151}$ Department of Physics, Technion: Israel Institute of Technology, Haifa, Israel\\
$^{152}$ Raymond and Beverly Sackler School of Physics and Astronomy, Tel Aviv University, Tel Aviv, Israel\\
% ...
$^{157}$ Department of Physics, University of Toronto, Toronto ON, Canada\\
% ...
\end{flushleft}


\clearpage

\end{document}